\documentclass{mn2e}

\usepackage{graphicx}
\usepackage{txfonts}

\begin{document}

\title[High Precision Astrometry with E-ELT/MICADO]{High Precision Astrometry with MICADO at the European Extremely Large Telescope}

\author[S. Trippe et al.]{S. Trippe,$^1$\thanks{E-mail: trippe@iram.fr} R. Davies,$^2$ F. Eisenhauer,$^2$ N.M. F\"orster Schreiber,$^2$ T.K. Fritz,$^2$ R. Genzel$^{2,3}$ \\
$^1$ Institut de Radioastronomie Millim\'etrique, 300 rue de la Piscine, F-38406 Saint Martin d'H\`eres, France \\
$^2$ Max-Planck-Institut f\"ur extraterrestrische Physik, Giessenbachstrasse 1, D-85748 Garching, Germany \\
$^3$ Department of Physics, Le Conte Hall, University of California, CA 94720 Berkeley, USA
}

\date{Accepted 2009 October 26. Received 2009 October 26; in original form 2009 September 11.}

\pagerange{\pageref{firstpage}--\pageref{lastpage}} \pubyear{2009}

\maketitle

\label{firstpage}

\begin{abstract}
In this article we identify and discuss various statistical and systematic effects influencing the astrometric accuracy achievable with MICADO, the near-infrared imaging camera proposed for the 42-metre European Extremely Large Telescope (E-ELT). These effects are instrumental (e.g. geometric distortion), atmospheric (e.g. chromatic differential refraction), and astronomical (reference source selection). We find that there are several phenomena having impact on $\sim$100$\mu$as scales, meaning they can be substantially larger than the theoretical statistical astrometric accuracy of an optical/NIR 42m-telescope. Depending on type, these effects need to be controlled via dedicated instrumental design properties or via dedicated calibration procedures. We conclude that if this is done properly, astrometric accuracies of 40$\mu$as or better -- with 40$\mu$as/yr in proper motions corresponding to $\approx$20~km/s at 100~kpc distance -- can be achieved in one epoch of actual observations.
\end{abstract}


\section{Introduction}

The future optical/near-infrared European Extremely Large Telescope (E-ELT; see, e.g., Gilmozzi \& Spyromilio \cite{gilmozzi2008}), which is designed with a 42-metre aperture, will offer a substantial improvement in angular resolution compared to existing facilities. At wavelengths $\lambda=2\mu$m, diffraction-limited resolutions of  $\Theta\simeq10$mas will be achieved. In terms of angular resolution in the near infrared, the E-ELT will outperform existing 8--10m-class telescopes like the VLT or Keck by factors of $\approx$4--5 and the future James Webb Space Telescope (JWST) by factors of $\approx$7. This increase in angular resolution should translate into a corresponding improvement in astrometric accuracy.

In order to exploit the E-ELT's resolution, a German-Dutch-Italian-French consortium\footnote{The MICADO collaboration includes: MPE Garching, Germany; USM Munich, Germany; MPIA Heidelberg, Germany; NOVA (a collaboration of the universities of Leiden, Groningen, and ASTRON Dwingeloo), The Netherlands; OAPD Padova (INAF), Italy; LESIA Paris, France} proposed the Multi-AO Imaging Camera for Deep Observations (MICADO) in February 2008. As the spatial resolution of any ground-based observatory is initially limited by the atmospheric seeing, MICADO will be equipped with a multi-conjugate adaptive optics (MCAO) system for achieving the diffraction limit of the 42m-telescope. This system uses three natural and six laser guide stars for correcting the atmospheric turbulence in a wide ($>2'$) field of view  (Diolaiti et al. \cite{diolaiti2008}). Images will be recorded by an array of 4$\times$4 near-infrared (NIR) HAWAII-4RG detectors with 4096$\times$4096 pixels each, covering a FOV of 53'. The instrument is sensitive to the wavelength range $0.8-2.5\mu$m, thus covering the I, Y, J, H, K bands. For astrometric experiments the use of the data analysis software {\sl Astro-WISE} (Valentijn et al. \cite{valentijn2007}) is foreseen.

In order to achieve its science goals (see Sect.~2 for details), MICADO needs to reach a stable (time scales of years) astrometric accuracy of approximately 50$\mu$as. At present 8--10m class telescopes, accuracies of $\approx$0.5\% of a resolution element can be reached regularly (e.g. Fritz et al. \cite{fritz2009b}). Therefore from simple scaling of results our goal \emph{a priori} appears reasonable. However, at levels of the order of 100$\mu$as there are several sources of statistical and systematic errors which need to be taken into account carefully. In this article we discuss those effects and analyse strategies to bypass them. We conclude that reaching an astrometric accuracy of better than 50$\mu$as is highly challenging in terms of instrument design and data calibration but feasible.

MICADO's astrometric performance should be of the same magnitude as that of the future astrometry space mission GAIA (e.g. Jordan \cite{jordan2008}). GAIA will achieve accuracies better than $\approx$50$\mu$as only for bright (V$<$15.5) targets and only at the end of its mission. MICADO is expected to achieve this accuracy for targets with $K_{\rm AB}<26$. Other space missions like SIM PlanetQuest (e.g. Edberg et al. \cite{edberg2007}) or JASMINE (e.g. Gouda et al. \cite{gouda2007}) also aim specifically at bright targets in order to reach accuracies of $\approx$10$\mu$as (at best).

\begin{figure}
\includegraphics[height=8.8cm]{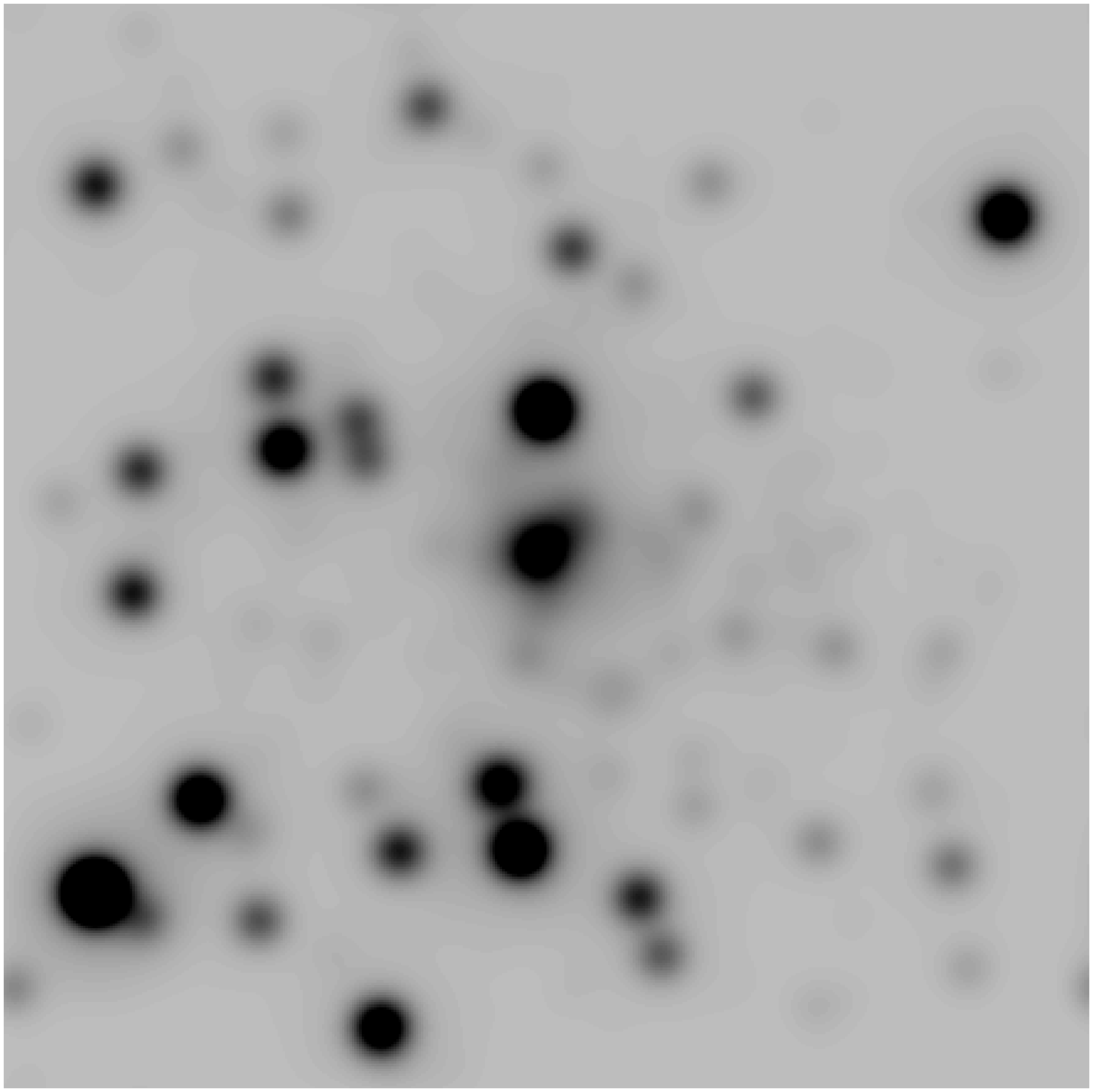}
\includegraphics[height=8.8cm]{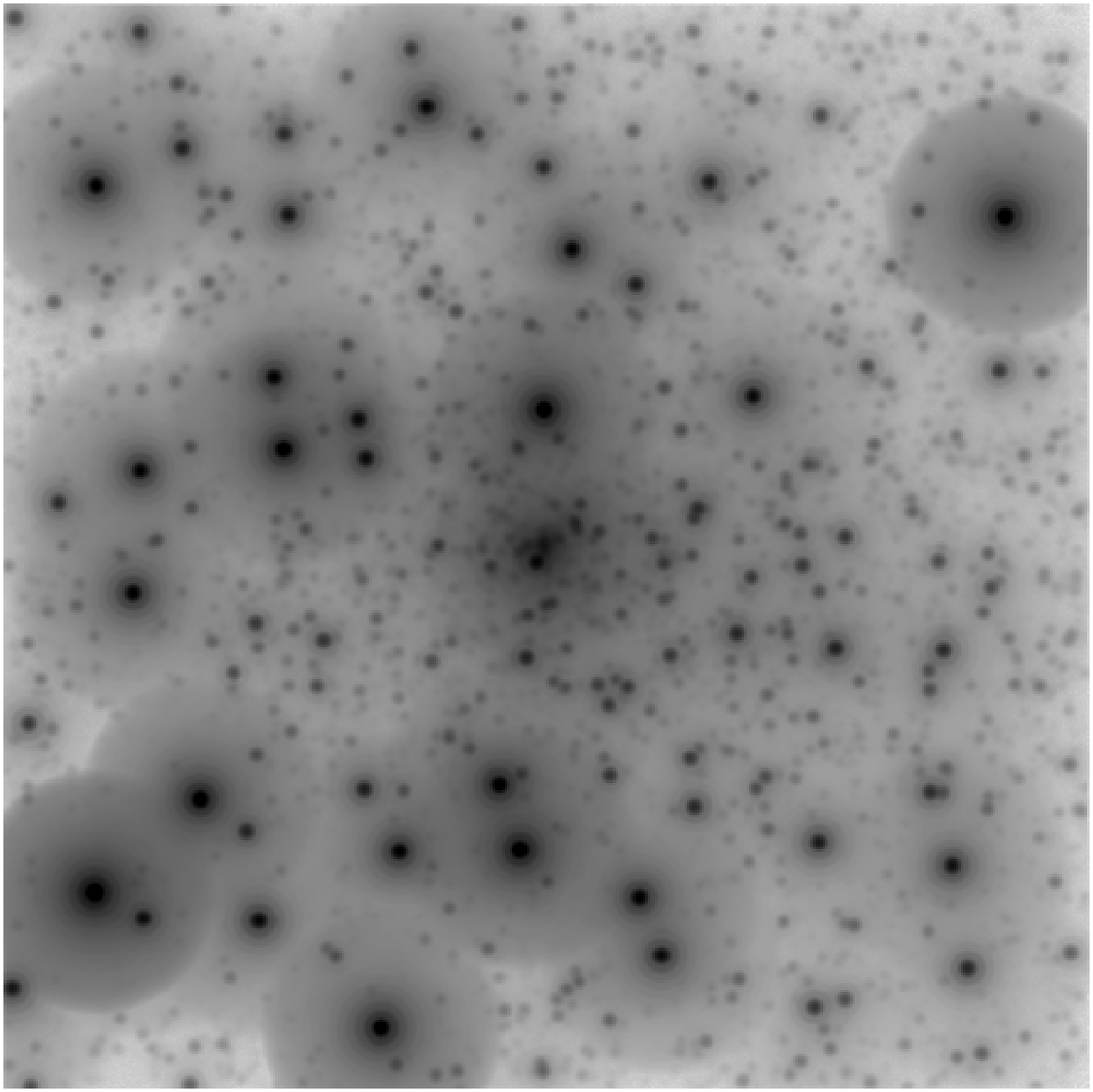}
\caption{An illustration of the expected performance of the E-ELT/MICADO system. These simulated maps show the central $1''\times1''$ (i.e. 8000AU$\times$8000AU) of the nuclear star cluster of the Milky Way at 2.2$\mu$m. \emph{Top panel}: The target region as observed with present day 8-10m class telescopes. The diffraction-limited resolution is $\approx$50mas. For comparison with actual observations, see, e.g., Genzel et al. (2003), Ghez et al. (2005). \emph{Bottom panel}: The same field as seen by MICADO. The angular resolution is $\approx$10mas. The improvement in detail and depth is obvious.}
\label{fig_eelt_gc}
\end{figure}

For illustration purposes, Fig.~\ref{fig_eelt_gc} shows simulated observations of the nuclear star cluster of the Milky Way using both present day 8-10m class telescopes and E-ELT/MICADO. Physical parameters of the star cluster (stellar density profile, luminosity function) are taken from Genzel et al. \cite{genzel2003}. We discuss technical details of our simulations in Sect. 4.1. These maps demonstrate the impressive progress to be expected with MICADO.

Although this study is set up for the specific case of MICADO, most of its results are valid in general and therefore of interest beyond the E-ELT community.

This paper is organised as follows. In Section~2, we discuss the science cases identified for MICADO. In Section~3, we review the concepts and techniques of accurate astrometry. In Section~4, we identify and analyse sources of systematic errors one by one and describe methods for minimizing those errors. We provide a summary of our results and an overall error budget in Section~5 and present our conclusions in Section~6.

\section{Science Cases}

As part of the instrument design study, the MICADO collaboration has identified and analysed (Renzini et al. \cite{renzini2008}, and references therein) several science cases for which the high astrometric accuracy of E-ELT/MICADO is crucial and promises major discoveries. We discuss them in the following one-by-one.

\subsection{Galactic Centers}

Located at a distance of $\approx$8~kpc, the nuclear region of the Milky Way is the closest galactic nucleus, hosting the supermassive black hole ($M_{\bullet}\approx4\times10^6M_{\odot}$) Sgr~A* (e.g. Gillessen et al. \cite{gillessen2009}). It is therefore a unique laboratory for exploring the regime of strong gravity, accretion onto black holes, and the co-evolution of dense star clusters and active galactic nuclei.

Present-day NIR instrumentation, e.g. VLT/NACO, provides astrometric accuracies down to $\approx$0.3~mas and angular resolutions down to $\approx$50~mas (e.g. Fritz et al. \cite{fritz2009b}). This allowed to identify several stars on Keplerian orbits around Sgr~A* with orbital periods down to $\approx$15 years and pericenter distances as small as $\approx$100~AU ($\approx$12~mas on sky; e.g. Gillessen et al. \cite{gillessen2009}). It made possible to study in detail the kinematics and the composition of the nuclear star cluster in the gravitational potential of the central black hole. With E-ELT/MICADO one can expect to achieve sensitivities that are more than five magnitudes fainter than for VLT/NACO. Angular resolutions and astrometric accuracies should also improve by factors of about five, meaning that proper motions of order 10$\mu$as/yr (400~m/s) can be detected within few years of observations. Such instrumental performance is necessary in order to adress several new questions (e.g. Gillessen et al. \cite{gillessen2009}):

\begin{itemize}

\item Identification of stars on closeby Keplerian orbits with periods of few years.

\item Measuring the \emph{prograde} relativistic orbit precession and testing other effects of general relativity.

\item Probing possible \emph{retrograde} orbit precession due to an extended mass component built from compact stellar remnants.

\item Analyzing the various separate kinematic structures of the nuclear star cluster, and searching for new ones.

\item Quantifying the binary star fraction in the nuclear cluster.

\end{itemize}

With MICADO, this type of analysis can be extended to other nearby galaxies. One obvious example is the core of M31 which hosts a $M_{\bullet}\approx1.4\times10^8_{\odot}$ black hole. Similar to the case of the Milky Way, M31's nucleus shows several distinct stellar populations: a triple nucleus and two nested star disks around the central black hole have been identified (Bender et al. \cite{bender2005}). Although M31 is more distant from earth by a factor $\approx$100 compared to the Galactic center, the larger mass of its black hole (by a factor of $\approx$35) causes stellar proper motions of about 6\% compared to those in the nuclear cluster of the Milky Way. Therefore kinematic analyses analoguous to the Galactic center experiment will require capabilities as predicted for E-ELT/MICADO.

Another example is Centaurus~A which hosts a $M_{\bullet}\approx5\times10^7_{\odot}$ black hole (Neumayer et al. \cite{neumayer2007}). Given its distance of $\approx$5~Mpc, proper motions of $\approx$10$\mu$as/yr correspond to $\approx$200~km/s. Thus measuring the motions of stars in the vicinity of the black hole is possible. Other galaxies might be interesting targets as well.

\subsection{Intermediate Mass Black Holes}

The expected high astrometric accuracy of MICADO opens a new window in the search for and analysis of intermediate mass black holes (IMBH), objects with masses of few thousand solar masses. In the last years, the detection of those objects in the Arches cluster (Portegies Zwart et al. \cite{portegies2006}), the star association GCIRS13 in the Galactic center (e.g. Maillard et al. \cite{maillard2004}), $\omega$~Cen (Noyola et al. \cite{noyola2008}), and other locations has been claimed. Most of these analyses are based on radial velocity dispersion profiles. This introduces systematic ambiguities as anisotropic velocity dispersions can mimick the presence of central point masses; this one can see for example in the anisotropy term in the Jeans equation. Therefore reliable (non)detections of IMBHs require measurements of stellar proper motions (e.g. Anderson \& van der Marel \cite{anderson2009}). Typical velocity dispersions $\sigma_*$ of star clusters are of the order 10~km/s. This corresponds to $\approx$50$\mu$as/yr at a distance of 40~kpc, meaning that for most of the Galactic star clusters a proper motion analysis is feasible with MICADO only. This allows

\begin{itemize}

\item contraining black hole masses in Galactic star clusters within few years,

\item probing the low-mass end of the $M_{\bullet}-\sigma_*$ relation,

\item testing the dynamical evolution of star clusters.

\end{itemize}

\subsection{Globular Clusters}

As discussed above, the astrometric accuracy of MICADO should allow measuring stellar proper motions of few km/s for most Galactic globular clusters. Additionally, direct measurements of cluster parallaxes become possible: for a distance of 40~kpc, the full parallax displacement is 50$\mu$as, corresponding to MICADO's predicted astrometric accuracy. This allows adressing several topics:

\begin{itemize}

\item The spatial distribution of globular clusters.

\item Cluster proper motions and their orbits around the Milky Way (e.g. Bedin et al. \cite{bedin2003,bedin2006}).

\item Internal cluster kinematics, including rotation.

\item Separating cluster members from field stars, thus making analyses of cluster star populations more reliable (see also Anderson et al. \cite{anderson2006}).

\end{itemize}

\subsection{Dark Matter In Dwarf Spheroidal Galaxies}

Cold dark matter models predict high mass densities and cuspy density profiles for the central regions of galaxy halos. In contrast, warm dark matter models predict substantially lower central densities and constant density cores at small radii. The dwarf spheroidal satellite galaxies of the Milky Way provide a unique laboratory to test those models. Their proximity of $\approx$100~kpc makes it possible to resolve individual stars and analyze their dynamics in the gravitational potentials of their galaxies. Present day studies are usually based on line-of-sight velocity dispersion profiles. However, there are degeneracies of velocity dispersion anisotropies with mass density profiles. Therefore any conclusive analysis requires measuring all three components of the velocity vectors of the tracer stars (e.g. Strigari et al. \cite{strigari2007}).

This type of studies requires accurate proper motion measurements with uncertainties of few km/s or better. MICADO will be able to provide accuracies of $\approx$5~km/s within few years of observations for targets about 100~kpc away. This makes MICADO a decisive tool for testing the validity of present-day dark matter models.

\section{The Astrometry Problem}

Throughout this paper, we use the term ``astrometry'' the following way. We discuss time-resolved \emph{relative} positions, meaning the positions of a science target with respect to a set of reference sources. Science target and reference sources are located in the field of view (FOV) of the camera, i.e. they are recorded simultaneously in the same science image.

\emph{Intra-}epoch measurements cover timelines that are so short that intrinsic motions of science targets or reference sources cannot be detected. This can be a set of images taken within the same night or a few adjacent nights. Any variations in measured positions are due to measurement errors and can be used to determine the position accuracies. All information obtained from this data set (i.e. images, coordinates, ...) can be combined, e.g. for improving the signal-to-noise ratio (SNR).

\emph{Inter-}epoch measurements cover timelines sufficient for detecting source motions. Those data cannot be combined in a straight forward manner. Intrinsic source motions and measurement errors interfere and need to be disentangled, usually meaning that calculating errors requires additional information. Given MICADO's proper motion accuracies of $\approx$50$\mu$as/yr or better (depending on the duration of the experiments), corresponding to $\approx$10~km/s at 40~kpc distance, intrinsic motion of Galactic stars is detected easily. Therefore it will be necessary to use extragalactic sources, including high redshift objects like QSOs, as references for some science cases.

If a set of science images is at hand, the general analysis recipe is as follows.

\emph{Step 1}. From each image $n$, one extracts the detector positions $\{{\bf X}^n\}$ of all sources of interest (science targets and reference sources). Detector positions need to be measured with high accuracies of order few milli-pixels (mpix). Existing centroiding, source profile fitting, and point spread function (PSF) correlation algorithms provide such accuracies (e.g. Diolaiti et al. \cite{diolaiti2000}; Berry \& Burnell \cite{berry2000}; Trippe \cite{trippe2008a}).

\emph{Step 2}. For each image $n$, the detector positions $\{{\bf X}^n\}$ need to be converted into global astrometric coordinates $\{{\bf x}^n\}$. The reference frame can be the detector coordinates of a selected zero-point image or any more general astrometric coordinate system. Using the detector positions of the reference sources $\{{\bf X}_{\rm ref}^n\}$ and their astrometric positions $\{{\bf x}_{\rm ref}^n\}$, one calculates a transformation

\begin{equation}
T_n: \{{\bf X}_{\rm ref}^n\} \longrightarrow \{{\bf x}_{\rm ref}^n\} ~.
\label{eq_trafo_get}
\end{equation}

\noindent
Obtaining the positions $\{{\bf x}_{\rm ref}^n\}$ requires some prior knowledge on the reference sources. One powerful approach is cross-calibration with other datasets, maybe from other wavelength regimes. A nice example is given by Reid et al. \cite{reid2007}. They use precise VLBI astrometry of SiO maser stars in order to define an astrometric reference frame in K-band images of the Galactic center.

If the reference sources are known to be not moving (e.g. extragalactic objects), one can set $\{{\bf x}_{\rm ref}^n\}=\{{\bf X}^0\}$; the index 0 indicates the selected zero-point in time. The same can be done if the reference source ensemble is (or is defined as being) at rest \emph{in average}, i.e. $\langle \{{\bf x}_{\rm ref}^0\} \rangle = \langle \{{\bf x}_{\rm ref}^n\} \rangle$ (e.g. a sufficiently large set of stars in a star cluster). In this case however, one will loose information on a global motion (drift, rotation, contraction, etc.) of the combined system ``science target $+$ reference sources''. In any case, the transformation $T_n$ is used to compute global astrometric coordinates for the science targets like

\begin{equation}
\{{\bf X}^n\} \longrightarrow T_n\left( \{{\bf X}^n\}\right) = \{{\bf x}^n\} ~.
\label{eq_trafo_apply}
\end{equation}

\noindent
Commonly, low-order ($<$5) 2-dimensional polynomial coordinate transformations

\begin{eqnarray}
x' = a_0 + a_1 x + a_2 y + a_3 x^2 + a_4 y^2 + a_5 x y + ... \\
y' = b_0 + b_1 x + b_2 y + b_3 x^2 + b_4 y^2 + b_5 x y + ...
\label{eq_trafo_poly}
\end{eqnarray}

\noindent
are used. In case prior knowledge on the geometry of the required transformation is available, one can use models with smaller numbers of free parameters (e.g. Montenbruck \& Pfleger \cite{montenbruck1989}; Anderson et al. \cite{anderson2006}; Trippe et al. \cite{trippe2008b}).

The number of available reference sources governs the maximum order of coordinate transformations (see step 2). A 1st order polynomial transform with six parameters requires three reference sources, i.e. $2\times3=6$ coordinates. A 2nd-order transform (12 parameters) requires six reference sources, and so on. Throughout this paper we assume that \emph{any two science images need to be connected via full astrometric transformations}. This means that we regard more simple methods of data combination like stacking, simple-shift-and-add etc. as non-astrometric and thus not usable for our purpose.

In order to judge the astrometric accuracy achievable with a system like MICADO, one has to distinguish statistical and systematic influences. The \emph{statistical} measurement accuracy is given by

\begin{equation}
\sigma_L = \frac{\lambda}{\pi D}\frac{1}{SNR} = 284\mu{\rm as}\left( \frac{\lambda}{2.17\mu{\rm m}}\right)\left( \frac{5{\rm m}}{D}\right)\left( \frac{100}{SNR}\right)
\label{eq_accuracy}
\end{equation}

\noindent
Here $\lambda$ is the wavelength, $D$ the telescope aperture, and $SNR$ the signal-to-noise ratio (Lindegren \cite{lindegren1978}). To give an example for the case of the E-ELT: with $\lambda=2.2\mu$m (K-band), $D=42$m, and $SNR=100$, one obtains a statistical astrometric accuracy $\sigma_L=34\mu$as. From this we see that -- in principle -- astrometric accuracies of $\sim$10$\mu$as can be obtained with the E-ELT. This means that any source of additional, especially \emph{systematic} error needs to be compensated down to this level if one actually wants to fully exploit the E-ELT's capabilities.

\section{Error Sources}

In total, we have identified ten effects that might have the potential to reduce the expected astrometric accuracy of MICADO substantially. We will discuss these ``Terrible Ten'' in the following subsections. The first three phenomena we analyze are instrumental, the next five are atmospheric, and the last two are astronomical.

\subsection{Sampling and Pixel Scales}

The detector position of a point source can be computed only if the source PSF is sufficiently sampled. If the pixel scale -- expressed in angular units per pixel -- is too large (undersampling), position information is lost because there is no unique mathematical description for the PSF profile anymore. Especially, a PSF can then be modelled by profiles with different centers of light (i.e. different detector positions). This effect is known as the \emph{pixel phase error}. It can reach magnitudes of several tenths of a pixel, thus providing an important boundary condition for the instrument design. A detailed description of this phenomenon is provided by Anderson \& King \cite{anderson2000}.

In order to identify the critical pixel scale of MICADO, we first created artificial PSFs for each of the bands I, J, H, and K. We modelled each PSF $P({\bf x})$ as a superposition of a 2-dimensional Airy function $A({\bf x})$ and a 2D Moffat profile $M({\bf x})$ like

\begin{equation}
P({\bf x}) = a A({\bf x}) + (1-a) M({\bf x}) ~.
\label{eq_psf}
\end{equation}

\noindent
For each PSF, a stochastic optimization routine\footnote{Implemented in the MPE data processing software {\sl DPUSER} developed by Thomas Ott; see {\tt http://www.mpe.mpg.de/$\sim$ott/\\dpuser/history.html}} adjusted the parameter $a\in [0,1]$ such that the resulting PSF profile showed the proper Strehl ratio. We took Strehl ratio estimates for MICADO from preliminary E-ELT adaptive optics (AO) system simulations (Liske \cite{liske2008}; M. Kissler-Patig \emph{priv. comm.}), the values are shown in Table~\ref{tab_strehls}.

\begin{table} 
\centering
\caption{Strehl ratio estimates as obtained from preliminary simulations (Liske 2008; M. Kissler-Patig \emph{priv. comm.}). The $\lambda_{\rm center}$ are the central wavelengths of the filters.}
\begin{tabular}{l c c c c}
\hline\hline
Band & I & J & H & K \\
\hline
$\lambda_{\rm center}$ [$\mu$m] & 0.900 & 1.215 & 1.654 & 2.179 \\
\hline
Strehl ratio [\%] & 2 & 18 & 35 & 53 \\
\hline
\end{tabular}
\label{tab_strehls}
\end{table}

For each of the four wavelength bands we examined pixel scales from 1 to 7~mas/pix in steps of 0.1~mas/pix. In each configuration, we placed the PSF at a random detector position and re-binned it to the corresponding pixel scale. After this, we fit the detector position with a 2D Gaussian light distribution. We iterated this procedure 250 times for each configuration. Pixel phase errors were the rms values of the distributions of the differences between true and measured positions.

\begin{figure} 
\includegraphics[height=8.8cm,angle=-90]{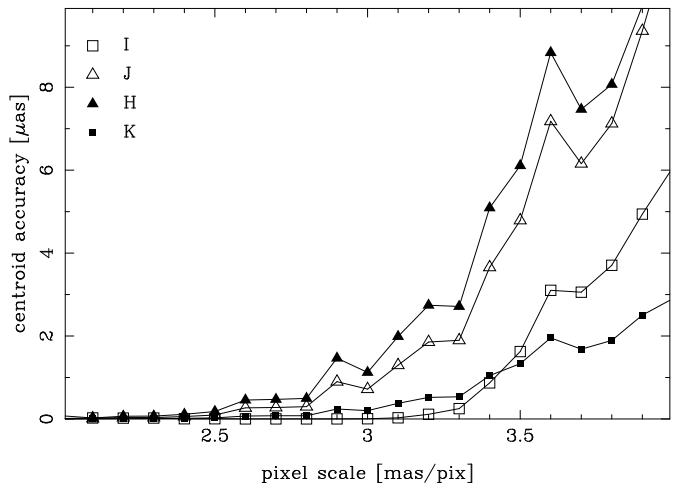}
\caption{Pixel phase error vs. pixel scale for I, J, H, K bands for \emph{isolated} sources. Model PSFs are superpositions of Airy and Moffat profiles with Strehl ratios as given in Table~\ref{tab_strehls}. For all bands, the errors are below $\approx$1$\mu$as for pixel scales smaller than 3~mas/pix.}
\label{fig_pix_isol}
\end{figure}

The results of our analysis are shown in Fig.~\ref{fig_pix_isol}. For clarity, we restrict the diagram to the decisive pixel scale range from 2 to 4~mas/pix. For all wavelength bands the errors are smaller than $\approx$1$\mu$as for pixel scales below 3~mas/pix. Given that the filter bands span a factor of 2.4 in wavelength and thus in $\lambda/D$, it might not be obvious why the critical pixel scales we find are that similar for all bands. This effect is caused by the differences in Strehl ratios: whereas the diffraction limited profile width decreases with decreasing wavelength, the Strehl ratio decreases, too, meaning a stronger atmospheric blur. In our case, the two effects roughly counterbalance each other; we actually find the smallest beam size for the H band PSF due to its combination of high SNR ($\approx$35\%) and small diffraction limit ($\lambda/D\approx8.1$~mas). This makes it possible to quote a single critical pixel scale for all bands. We therefore conclude that for \emph{isolated} point sources pixel scales up to 3~mas/pix can be used for MICADO without introducing noticable ($\approx$1$\mu$as) pixel phase errors. Therefore we use this scale as the standard value for the MICADO design.

In case of \emph{crowding}, the astrometric accuracy is additionally limited by source overlap. For this reason the MICADO design foresees a ``small scale  mode''. As the camera design is catoptric and all mirrors are fixed, we use a pick-up arm to image a small (6'', using one out of the 4$\times$4 detectors) FOV with a reduced pixel scale of 1.5~mas/pix. The reduced scale helps to better separate close by sources and such reduce the misplacements of the source centroids.

In order to quantify this effect, we examined our two pixel scales, 1.5~mas/pix and 3~mas/pix, for each of the bands H and K, meaning four configurations in total. For each configuration we simulated a crowded star field by placing PSFs at random positions. We used the same model PSFs as for the case of isolated sources. We applied a luminosity profile corresponding to the K-band luminosity function of the Galactic bulge which is $d\log n / d\log S \approx -0.8$ (Zoccali et al. \cite{zoccali2003}; $n$ is the source number, $S$ is the flux). The dynamic range of the star sample was 10 magnitudes, i.e. a factor of 10,000 in flux. Source densities were $\approx$3000 stars per square arcsecond. We also used these routines (with modified parameters) to create the map shown in Fig.~\ref{fig_eelt_gc}.

In each simulated map, we searched for stars and calculated their detector positions using the PSF fitting routine {\sl StarFinder} (Diolaiti et al. \cite{diolaiti2000}). We derived median uncertainties vs. fluxes from the distributions of the differences between true and measured positions. As the absolute values of these errors are functions of several parameters like fluxes, luminosity profiles, and source densities, we converted our results into relative numbers, using the errors obtained for the 3~mas/pix scales as references.

\begin{figure} 
\includegraphics[height=8.8cm,angle=-90]{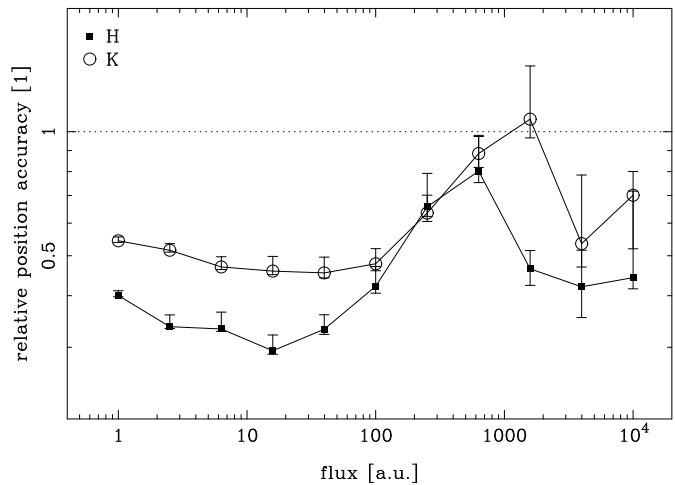}
\caption{Relative median astrometric error vs. source flux for \emph{crowded} sources, separately for H and K bands. The results for the pixel scales 1.5~mas/pix are given as fractions of the uncertainties found for 3~mas/pix. The horizontal dashed line indicates a ratio of 1. Error bars indicate the 68\% uncertainty ranges. This analysis shows that reducing the pixel scale improves the typical accuracies by factors $\approx$2--3.}
\label{fig_pix_crowd}
\end{figure}

The outcome of our analysis is presented in Fig.~\ref{fig_pix_crowd}. Reducing the pixel scales from 3~mas/pix to 1.5~mas/pix improves the typical accuracies by factors of $\approx$2--3. The effect is stronger in H than in K band; this indicates that images with smaller beam sizes profit more from a small pixel scale. For the brightest sources (more than $\approx$500 units) the differences between the pixel scales are (at least in K band) not very pronounced any more because (a) the number of very bright sources is small and (b) they outshine most of their neighbours. We can thus conclude that for \emph{crowded} fields the use of small pixel scales down to 1.5~mas/pix is indeed important for keeping a high level of astrometric accuracy.

Combining all results for isolated and crowded sources, we can conclude the following for the design and the operation of MICADO:

\begin{itemize}

\item  For sufficiently \emph{isolated} target sources, pixel scales smaller than or equal 3~mas/pix are free of noticable ($\approx$1$\mu$as) systematic uncertainties and suited for accurate astrometry. This scale is therefore going to be the standard pixel scale for MICADO.

\item  For the special case of \emph{crowded} sources, smaller pixel scales down to 1~mas/pix can substantially (factors $\approx$2) reduce the astrometric errors introduced by source overlap. We therefore implemented a ``small-scale mode'' with a reduced pixel scale of 1.5~mas/pix into the MICADO design. This mode will be used to map very crowded regions.

\end{itemize}

\subsection{Instrumental Distortion}

The geometric distortion of an optical system can seriously limit its astrometric accuracy. In the following, we discuss \emph{non-linear} distortions. \emph{Linear} terms like shifts, rotations, scalings, or shear, are absorbed by 1st (or higher) order coordinate transforms (see Sect.~3). In case of non-linear distortion, the effective pixel scale is a function of detector position (e.g. Greason et al. \cite{greason1994}, and references therein). 

For MICADO, the amount of distortion to be expected (meaning the difference between imaged and theoretical image positions) is of the order of few per cent. The largest numbers, about 5\%, we find for the case of an Offner design for the camera. Although other optical designs might have the potential to reduce the amount of distortion down to $\approx$0.3\% (Dierickx~\cite{dierickx2008}), it cannot be neglected. Across a FOV of $1'$, a distortion of 0.3\% corresponds to a misplacement of $0.18''$. For comparison: an astrometric accuracy of $\approx$50$\mu$as across the same FOV corresponds to a relative error of less than $10^{-6}$. Such low uncertainties cannot be achieved by design. In order to reach the desired astrometric accuracies, the distortion must be compensated by dedicated calibration schemes -- regardless of its amplitude. 

Another effect to be considered is imperfect fabrication of the detectors. The HAWAII-4RG detectors to be used for MICADO are designed with pixels sizes of $15\mu$m. The position accuracy of these pixels in the detector grid is $\approx$10nm; for a pixel scale of 3~mas/pix this corresponds to an astrometric uncertainty of $\approx2\mu$as (Richard Blank\footnote{Teledyne Imaging Sensors, Camarillo, USA} \emph{priv. comm}), i.e. is hardly noticable. However, any distortion calibration scheme must provide the ability to catch potential inaccuracies of the detectors (see also Anderson \cite{anderson2002} for the case of HST/WFC).

Depending on the complexity of the distortion, two methods for its description are possible. \emph{Analytic} descriptions make use of analytic parametrizations analogous to the coordinate transforms discussed in Sect.~3. Most commonly, 2D polynomials up to about 5th order are used as distortion models (e.g. Greason et al. \cite{greason1994}, and references therein). This approach allows covering all effects up to a selected order without prior knowledge on the geometry of the problem. The disadvantage of this method is the large number of model parameters to be calculated. A recent example for this approach is the analysis of the nuclear star cluster of the Milky Way by Ghez et al. \cite{ghez2008} and Lu et al. \cite{lu2009}. They use a polynomial model to correct the geometric distortion of the NIRC and NIRC2 cameras at the W.M. Keck Observatory.

If a physical model for the distortion is at hand, a corresponding model can be substantially simpler. A recent example for this approach is the analysis of the nuclear star cluster of the Milky Way by Trippe et al. \cite{trippe2008b} and Gillessen et al. (\cite{gillessen2009}). They use the 3rd-order model

\begin{equation}
{\bf r} = {\bf r'}(1-\beta {\bf r'}^2)
\label{eq_distortion}
\end{equation}

\noindent
with
$$
{\bf r} = {\bf x} - {\bf x}_C \hspace{1cm} {\rm and} \hspace{1cm} {\bf r'} = {\bf x'} - {\bf x}_C
$$
\noindent
(e.g. J\"ahne \cite{jaehne2005})\footnote{See also the electronic manual of the public Gemini North Galactic Center Demonstration Science Data Set for another application on Galactic Center imaging data.} in order to correct the distortion of the imager NAOS/CONICA at the VLT. Here ${\bf x}$ and ${\bf x'}$ are the true and distorted image coordinates respectively, $\beta$ is a parameter describing the strength of the grid curvature, and ${\bf x}_C\equiv(x_C,y_C)$ is the zero point of the distortion on the detector. This approach has the obvious advantage that it requires only three parameters ($\beta,x_C,y_C$) for a 3rd-order distortion model compared to 20 for the case of a full polynomial solution. The most important disadvantage is the need for accurate \emph{a priori} knowledge of the distortion geometry.

In cases where analytic solutions are not feasible or not accurate enough, \emph{empiric} descriptions might be used instead or in addition. This means that the information of interest is stored in look-up tables. Those tables usually have the dimensions of the detector(s) and give the amount of distortion (or correction) for each pixel (separately for $x, y$). This approach is necessary if significant high-frequency distortion -- meaning that the spatial scales of the signal are small compared to the detector size -- is present. For the specific case of MICADO, the gaps between the detectors will introduce discontinuities into any astrometric solution. This effect could be caught by using lookup tables.

In a detailed analysis of the WFC instrument aboard the Hubble Space Telescope, Anderson \cite{anderson2002} uses a combined ``polynomial model plus lookup-table'' ansatz to model the distortion of the camera. He achieves residuals smaller than 0.01 pixels with this method. For the case of MICADO, this would mean residual errors below $\approx$30$\mu$as which is an amount acceptable for our purpose.

There are several methods to extract the distortion parameters. \emph{On-sky methods} make use of dedicated observations of astronomical targets; usually, star clusters are used (e.g., Anderson \cite{anderson2002}). If the true (astrometric) coordinates of the sources in the target field are known, one can derive the distortion parameters by comparing true and observed (detector) positions. If the true source positions are not known, one can observe the target field many times with slightly different telescope pointings. In this case, one compares the pairwise distances of objects that are present in two or more images. Modulations in these distances as function of detector position are equivalent to (non-linear) distortion.

By construction, on-sky methods are sensitive to distortions introduced by the atmosphere. Given the high accuracies we seek, those methods are not practicable (at least stand-alone) for the case of MICADO but will be a secondary approach.

\emph{In-lab methods} characterize the instrument with dedicated measurements in the laboratory or at the telescope. One example is the ``north-south test'' used for the spectro-imager SINFONI at the VLT (e.g. Abuter et al. \cite{abuter2006}). These methods use devices that illuminate the detectors with well-defined images or light patterns. Comparing the theoretical with the observed images allows for a description of the distortion.

Given that in our acuracy regime the atmosphere can severely limit the quality of on-sky calibration images (see Sect. 4.4 -- 4.8 for details), we assume that we need to implement an internal calibration device into MICADO. For this purpose, we examined the use of a calibration mask located in the focal plane of the imager. Such a mask could be a regular pattern of holes in an intransparent material. However, the following calculation shows that this is actually challenging. With MICADO, we want to observe a FOV with extension $x=1'$ with an accuracy $\delta x=10\mu as$, meaning a relative accuracy of
$$
\delta x/x \approx 1.7\times10^{-7} ~.
$$
The width of MICADO's focal plane will be approximately $l=0.25$m. This scales to a positioning accuracy for any reference located in the focal plane of
$$
\delta l = l\times\delta x/x \approx 4.2\times10^{-8} \rm m ~.
$$ 
As we see here, the positions of the holes in our calibration mask need be known with accuracies of about 40nm (the hole diameters would be $\approx$30$\mu$m corresponding to diffraction-limited point sources). This result does not mean that the fabrication process needs to be accurate at this level. Instead, it is sufficient to map the hole positions with accuracies of $\approx$40nm after the making of the mask. State-of-the-art photolithographic techniques provide the required production and mapping accuracies at standard temperatures ($\approx$300K). With this approach, the calibration mask would be a transparent quartz plate with a chrome cover into which the holes are etched. Although such a system is thermally very stable, the impact of possible non-linear thermal deformation needs to be investigated via dedicated laboratory experiments (B. Lorenz\footnote{Center for NanoSciences, Ludwig Maximilians University, Munich, Germany}, D. Rose\footnote{Rose Fotomasken, Bergisch Gladbach, Germany} \emph{priv. comm.}). We consider the calibration mask approach to be the easiest one in terms of telescope and instrument design. We have not yet investigated in detail more ``exotic'' ideas, e.g. illuminating the detectors with a well-defined diffraction pattern or spectrum.

As we see from the discussion in this section, it is crucial to gather the maximum amount of information on the instrumental distortion of MICADO. We therefore conclude that the following steps are necessary:

\begin{itemize}

\item  Estimating amount and geometrical structure of the distortion theoretically from the optics design.

\item  Careful mapping of the camera in the lab and at the telescope. Implementation of an internal calibration device, e.g. a calibration mask.

\item  Testing the system on-telescope for any evolution of the distortion. Evolution parameters can be time (aging effects), telescope orientation (gravitational flexure), etc.

\item  Additional regular dedicated on-sky calibration observations of sufficient astronomical targets, e.g. star clusters, as secondary tests.

\end{itemize}

\subsection{Telescope Instabilities}

Instabilities of the telescope system have the potential to affect astrometric experiments. For the present design of the E-ELT one expects relative intra-night plate scale variations of $\approx$0.1\% (Gonzalez \cite{gonzalez2008}). Across a FOV of 1', this corresponds to position variations of order 60~mas. Another effect are adapter-rotator instabilities that can introduce systematic frame-to-frame rotation. Those rotations introduce position misplacements of the same order of magnitude (see, e.g., Trippe et al. \cite{trippe2008b} for the case of VLT/NACO).

Fortunately, those effects -- shifts, global scalings, rotations -- are linear in geometry. Therefore they can be controlled via coordinate transforms of 1st or higher order without additional calibration steps. This statement does not hold however for gravitational flexure effects that introduce a time-variable non-linear instrumental distortion. This phenomenon is covered by our discussion of instrumental distortion in the previous subsection.

\subsection{Achromatic Differential Atmospheric Refraction}

Any ground-based position measurement is affected by atmospheric refraction. As we discuss relative position measurements of sources located in our FOV, we have to take into account any \emph{relative} or \emph{differential} atmospheric refraction. In this case, we have to discriminate achromatic and chromatic effects.

Achromatic differential refraction is caused by the slight difference in zenith angles $\zeta$ of two (or more) sources located within the same FOV. For each source, the atmospheric refraction leads to a deviation between physical and observed zenith angles. For two sources at slightly different zenith angles, those deviations will be different. Therefore the observed distance between the two targets deviates from the physical one and needs to be corrected. The required correction $\Delta x$ of the distance between two sources 0 and 1 is approximately given by the relation

\begin{equation}
\Delta x = (1+\tan^2\zeta_1)(A+3B\tan^2\zeta_1)\Delta\zeta ~.
\label{eq_refraction_achro}
\end{equation}

\noindent
Here $\zeta_1$ is the zenith angle of source 1, $\Delta\zeta$ is the \emph{observed} zenith separation, $A$ and $B$ are constants. Detailed calculations (Gubler \& Tytler \cite{gubler1998}) show that the linear terms of this effect are of the order of several milli-arcsec, whereas the quadratic terms are as small as $\approx$1$\mu$as. For example: with $\zeta_1=45^{\circ}$ and $\Delta\zeta=30''$, the correction amounts to $\approx$15~mas in the first-order terms and $\approx$2$\mu$as in second order. Fortunately, these terms can be absorbed by quadratic coordinate transforms (see Sect.~3). From this we conclude that we can neglect achromatic differential atmospheric refraction for the purpose of our analysis.

\subsection{Chromatic Differential Atmospheric Refraction}

Chromatic differential refraction (CDR) or \emph{atmospheric dispersion} is a more severe problem for accurate astrometry than the achromatic case. As the refractive index $n$ of the atmosphere is a function of wavelength, the observed angular distance between two sources is a function of the (relative) source colours. For a given \emph{true} zenith distance $\zeta_t$, the deviation from the \emph{apparent} zenith distance $\zeta_a$ (in radians) follows the approximative relation

\begin{equation}
\zeta_t - \zeta_a \simeq R\tan\zeta_t = \left(\frac{n^2-1}{2n}\right)\tan\zeta_t
\label{eq_refraction_chrom}
\end{equation}

\noindent
Here $R$ is the refraction constant. For standard conditions the refractive index is given by

\begin{equation}
(n-1)\times 10^6 = 64.328 + \frac{29498.1\times 10^{-6}}{146\times 10^{-6}-s^2} + \frac{255.4\times 10^{-6}}{41\times 10^{-6}-s^2}
\label{eq_refraction_index}
\end{equation}

\noindent
with $s=\lambda^{-1}$, $\lambda$ being the vacuum wavelength in nm (e.g. Cox \cite{cox2000}). Of course, this has impact on relative astrometry only if the two sources have different colours. For observations of two stars with broadband JHK filters, astrometric errors are of the order of 1~mas (within a wide range, depending on zenith angles, angular distances, and source colours). We therefore need to correct the CDR in order to meet our desired accuracies.

In the following, we investigate the use of an atmospheric dispersion corrector (ADC) placed into the optical path. This ADC is a pair of ZnSe/ZnS biprisms which refracts -- for a given zenith angle -- the infalling radiation such that the CDR is compensated. In order to use this element for a range of zenith angles, the biprisms can be rotated around the optical axis relative to each other; this controls the strength of refraction. Such a system was built and operated successfully for the SHARP~II+ NIR camera system (Eisenhauer \cite{eisenhauer1998}). The required optical components are available in the spatial dimensions needed for MICADO (S.~Koebele \emph{priv. comm.}). In order to decide on the design strategy, we explored three options in detail:

\begin{figure}
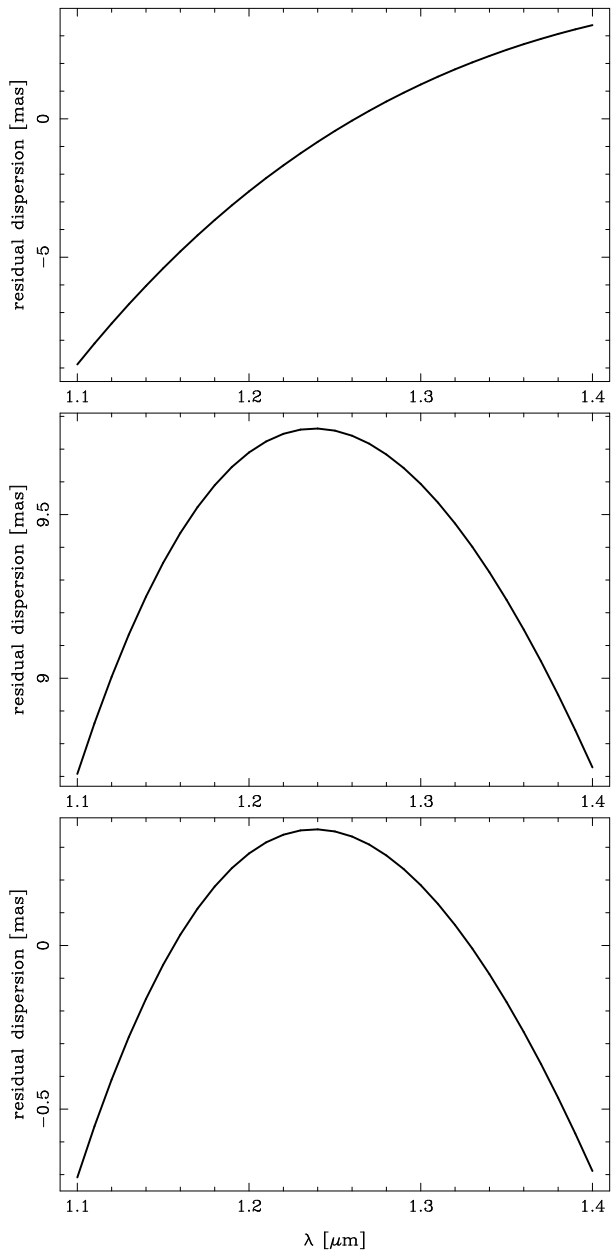
 
\includegraphics[height=8cm,angle=-90, trim = 0mm 0mm 12mm 0mm, clip]{f05a.eps}
\includegraphics[height=8cm,angle=-90, trim = 0mm 0mm 12mm 0mm, clip]{f05b.eps}
\includegraphics[height=8cm,angle=-90]{f05c.eps}
\caption{Residual dispersion vs. wavelength in J band after correction with an ADC assuming $\zeta=45^{\circ}$. \emph{Top panel}: Using one non-tunable ADC for the entire JHK band range. \emph{Center panel:} Using one tunable ADC for the entire JHK band range tuned to J band.  \emph{Bottom panel}: Using an ADC specifically designed for optimum correction in J band. Please note the changes in scales. The non-tunable ADC results in a very asymmetric correction curve. The other two curves are offset by about ten times their amplitudes from each other, but are very symmetric in their shapes.}
\label{fig_adc}
\end{figure}

\begin{enumerate}

\item The use of one non-tuneable ADC optimized for simultaneous correction of the full wavelength band range J, H, $\rm K_s$ ($\lambda=1.1...2.35\mu$m).

\item One ADC covering the entire band range that can be tuned in wavelength to bands J, H, and $\rm K_s$, respectively.

\item Three ADCs, one for each of the three bands J, H, and $\rm K_s$.

\end{enumerate}

\noindent
For each option (1, 2, 3), we computed residual dispersions (dispersions left after correction) vs. wavelength for each of the bands J, H, and $\rm K_s$. All calculations used zenith angles $\zeta=45^{\circ}$. Our computations made use of numerical optimization routines implemented into the software package {\sl Mathematica}\footnote{Wolfram Research, Inc., Champaign, IL, USA}. The results for J band are given in Fig.~\ref{fig_adc}. Whereas option 1 results in a very asymmetric curve of correction, options 2 and 3 are very similar in shape and highly symmetric. As we see here, one can reduce the residuals by using an ADC specifically designed for the wavelength band analyzed (option 3) by factors of about ten compared to a tunable ADC (option 2). However, the impact on astrometry is actually given by the \emph{relative} displacement of two sources with different colours. This means that the residual astrometric error is mainly controlled by the symmetry of the curves, not by their amplitudes or absolute levels. We may therefore expect that options 2 and 3 result in very similar astrometric accuracies.

For a quantitative description of the residual astrometric errors we analyzed two hypothetical science cases. We examined

\begin{itemize}

\item  an observation of two black bodies of very different temperatures ($T_1=$3,000~K, $T_2=$30,000~K), crudely corresponding to M5 and B0 main sequence stars, respectively (``case A''),

\item  the case of two black bodies at $T=$5,800~K (e.g. sun-like stars) affected by very different extinctions of $A_V=25$ and $A_V=35$, respectively (``case B''). For this case, we used the extinction law by Draine \cite{draine1989} as taken from Lutz et al. \cite{lutz1996}.

\end{itemize}

For each option, science case, and filter band we computed relative shifts in zenith angle between the sources 1 and 2 via

\begin{equation}
\Delta\zeta_{2-1} = \frac{\int \Delta\zeta(\lambda) S_{\lambda} d\lambda}{\int S_{\lambda} d\lambda} |_2 - \frac{\int \Delta\zeta(\lambda) S_{\lambda} d\lambda}{\int S_{\lambda} d\lambda} |_1 ~.
\label{eq_refraction_zshift}
\end{equation}

\noindent
We present the results of our analysis in Table~\ref{tab_adc}. As we see here, using a non-tuneable ADC (option~1) for the entire JHK band range leaves us with errors of roughly $\approx$100$\mu$as in H band which contains the center of correction. When going to $\rm K_s$ and J bands, the residuals increase to $\approx$200$\mu$as and $\approx$700$\mu$as, respectively. All numbers exceed any acceptable value by factors of about ten or more. This clearly rules out option~1.

\begin{table} 
\centering
\caption{Relative zenith angle shifts computed for the options, science cases, and filter bands discussed in the text. All numbers are in $\mu$as.}
\begin{tabular}{l c c c}
\hline\hline
Band & J & H & $\rm K_s$ \\
\hline
Option 1: &  &  &  \\
Case A & 730 & 73 & 196 \\
Case B & 642 & 112 & 250 \\
\hline
Option 2:  &  &  &  \\
Case A & 9 & 9 & 6 \\
Case B & 35 & 23 & 21 \\
\hline
Option 3:  &  &  &  \\
Case A & 9 & 9 & 2 \\
Case B & 35 & 22 & 11 \\
\hline
\end{tabular}
\label{tab_adc}
\end{table}

Using a tuneable ADC (option~2) or three ADCs optimized for the three bands (option~3) reduces the residual zenith angle displacements to $\approx$10$\mu$as. This result complies with our requirements and shows that our approach can indeed achieve the necessary accuracies for realistic science cases. As expected from Eqs.~\ref{eq_refraction_chrom} and \ref{eq_refraction_index}, the errors increase with increasing frequency. Therefore one should avoid observations at very short (shorter than J) wavelengths. Additionally, one should consider the use of narrow-band filters in case of (a) observations at short wavelengths or (b) extreme relative source colours.

Another calibration step one should consider is an \emph{a posteriori} correction. As we discuss above, the impact of CDR on astrometric solutions can be quantified analytically \emph{if} the relative source colours are known. However, this requires careful monitoring of the atmosphere (see, e.g., Helminiak~\cite{helminiak2009}).

From our analysis of chromatic differential atmospheric refraction, we conclude the following:

\begin{itemize}

\item  CDR distorts astrometric solutions by up to few mas depending on source colours. We need to implement a dedicated correction.

\item  Using a tuneable ZnS/ZnSe atmospheric dispersion corrector reduces the astrometric signal caused by CDR to $\approx$10$\mu$as in JHK bands asuming realistic science cases. We therefore highly recommend to implement such a device into MICADO.

\item In case of extreme relative source colours or observations at short (shorter than J-band) wavelengths, one should consider the use of narrowband filters additional to an ADC.

\item  If the source colours and the atmospheric conditions (temperatures, pressures, humidities) are known with sufficient accuracies, one might additionally apply an analytic \emph{a posteriori} correction of astrometric data.

\end{itemize}

\subsection{Guide Star Measurement Errors}

MICADO's MCAO system is designed to make use of three natural guide stars that are located (at a priori arbitrary positions) in the field of view. Additionally, the use of six laser guide stars is foreseen. The stars are observed simultaneously as references for correcting the wavefront deformation imposed on astronomical signals by the atmosphere. The natural guide stars are used for correcting low-order effects, whereas the laser guide stars correct high-order distortions. The LGS high-order correction does not make use of the guide star positons. 

This is different for the NGS low-order correction. This calibration step requires knowledge of the relative guide star positions. The guide star measurements will have finite errors due to atmospheric fluctuations that introduce a ``position wander''. Uncertainties in the measured guide star position will introduce time-variable distortion into the AO corrected FOV. For N stars, one may expect distortions up to order N$-$1, i.e. up to 2nd order for the case of three natural guide stars. These distortions will differ from image to image. Therefore it is necessary to combine images with full coordinate transform of minimum order two. If this is done, the effect will be compensated completely.

\subsection{Differential Tilt Jitter}

The light from a science target and the light from an adaptive optics system reference source travel through different columns of atmospheric turbulence. An AO system applies a tip-tilt correction to the signal received from the AO reference source. This correction is slightly different for other positions in the field of view. Therefore, any two objects in the observed field suffer from \emph{differential tilt jitter}: a random, achromatic, anisotropic fluctuation of the observed angular distance of the two sources (e.g., Britton~\cite{britton2006}; Cameron et al. \cite{cameron2009}). In first order, the rms of this fluctuation follows the relation

\begin{equation}
\sigma_{TJ} \propto \theta \times D^{-7/6} \times \left(\frac{\tau}{t}\right)^{1/2} ~.
\label{eq_tiltjitter}
\end{equation}

\noindent
Here $\theta$ is the angular distance between the sources, $D$ is the telecope aperture, $\tau$ is the aperture wind crossing time (approximately: $D$ divided by wind speed), and $t$ is the integration time. Tilt jitter is an anisotropic effect with

\begin{equation}
\sigma_{\parallel} = \sqrt{3}\sigma_{\perp} \approx 1.732\sigma_{\perp}
\label{eq_tiltjitter_aniso}
\end{equation}

\noindent
where $\sigma_{\parallel}$, $\sigma_{\perp}$ denote the tilt jitter rms parallel and perpendicular to the line connecting the two sources, respectively.

\begin{figure} 
\includegraphics[height=8.8cm,angle=-90]{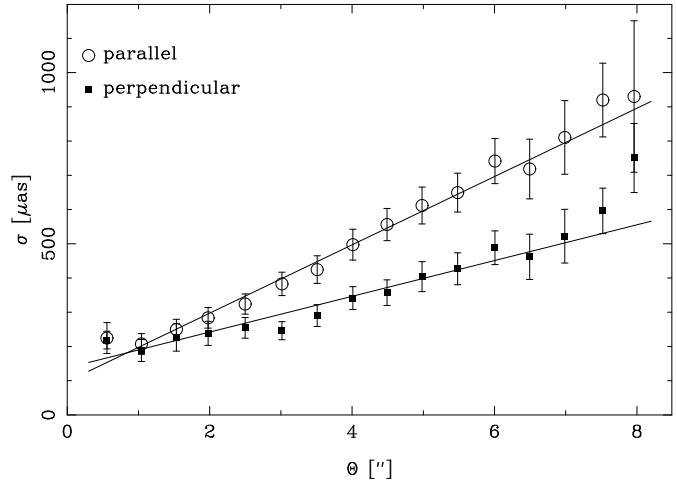}
\caption{A differential tilt jitter signal as observed in VLT images. This diagram shows the error on the measured distance between two stars $\sigma$ as a function of star angular distance $\theta$. Errors are calculated parallel ($\parallel$) and perpendicular ($\perp$) to the line connecting the two stars on sky. Points with errorbars are data, continuous lines indicate the best fitting linear models. From the line slopes one finds $\sigma_{\parallel}/\sigma_{\perp}=1.91\pm0.22$, i.e. $\sqrt{3}$ within errors as expected fom Eq.~\ref{eq_tiltjitter_aniso}. This result is taken from Fritz (2009).}
\label{fig_tilt}
\end{figure}

In Fig.~\ref{fig_tilt} we give an example for a tilt jitter signal observed in images taken with an 8-m-class telescope. This result is from the work by Fritz \cite{fritz2009} who analyzed diffraction-limited VLT images of the nuclear cluster of the Milky Way. For pairs of stars, he computed the uncertainties of the measured distances parallel and perpendicular to the lines conecting the two stars. Fig.~\ref{fig_tilt} shows the uncertainty as function of star angular distance. By means of linear fits one finds a relation $\sigma_{\parallel}/\sigma_{\perp}=1.91\pm0.22$, i.e. $\sqrt{3}$ within errors -- as expected fom Eq.~\ref{eq_tiltjitter_aniso}. From the line slopes and the known integration times one can estimate an aperture wind crossing time of $\tau\approx0.6$s. For details refer to Fritz \cite{fritz2009} and Fritz et al. \cite{fritz2009b}.

The impact of this effect on astrometry can be substantial. To give a reference: based on observations with the Hale 200 inch telescope, Cameron et al. \cite{cameron2009} find astrometric uncertainties of $\approx$75~mas for $D=5$m, $\theta=60''$, and $t=\tau=0.2$s (i.e. $\tau/t=1$). For a realistic intra-epoch E-ELT observation with $D=42$m and $t=100\tau\approx100$s (i.e. $\tau/t=0.01$), this scales to $\sigma_{TJ}\approx$600$\mu$as. Averaging out this error to reasonable scales (less than 50$\mu$as) would take about four hours. This imposes severe constraints on any astrometric observation.

The calculation given above is strictly valid only for single-conjugate AO (SCAO) systems, i.e. AO systems using one reference source. However, MICADO is designed to use an MCAO system with three natural guide stars initially. The use of multiple guide star should reduce the tilt jitter error substantially. Ellerbroek \cite{ellerbroek2007} finds that for a 30m telescope and the case of three guide stars arranged in an equilateral triangle the error is reduced by a factor $\approx$6 compared to the SCAO case. Scaling the results by Ellerbroek \cite{ellerbroek2007} to the case of the E-ELT, one finds 

\begin{equation}
\sigma_{TJ}\approx430\times t^{-1/2} \ \mu\rm as 
\label{eq_tilt_mcao}
\end{equation}

\noindent
with $t$ being the integration time in seconds and $\theta=60''$. This means that accuracies of $\approx$10$\mu$as can be achieved within integration times of about 30 minutes. 

A helpful property of tilt jitter is the fact that it is random in time, but correlated in space. This is also reported by Fritz \cite{fritz2009} who finds that already using linear coordinate transforms between images reduces the error by a factor $\approx$2. According to Cameron et al. \cite{cameron2009}, the use of coordinate reference frames based on weighted pairwise distances between a target source and several reference sources is able to catch the tilt jitter error. This approach is similar to the concept of kriging (e.g. Clark \& Harper \cite{clark2000}). It requires some tens of astrometric reference points in the FOV in order to obtain a sufficient number of pairwise baselines. Additionally, it requires any intra-epoch dataset to consist of some tens of individual exposures instead of few long-term integrations. This is necessary in order to estimate the uncertainty (and thus the weight) of each pairwise source distance from its histogram. If this scheme is applied, the error can be reduced to $\approx$10$\mu$as within few minutes of integration.

From our discussion on differential tilt jitter we conclude for MICADO:

\begin{itemize}

\item  Differential tilt jitter can introduce errors of order $\approx$100$\mu$as into typical E-ELT/MICADO observations. This effect is not a challenge for the instrument design but for the planning of observations and data calibration.

\item  The tilt jitter error integrates out with time as $\propto t^{-1/2}$. For the E-ELT, reducing the error to order of $\approx$10$\mu$as, integration times of at least $\approx$30~min per intra-epoch dataset will be necessary.

\item  The tilt jitter error can be calibrated out (down to $\approx$10$\mu$as) by using dedicated astrometric reference frames and transforms. This requires (a) some tens of reference sources in the FOV and (b) some tens of individual exposures per intra-epoch dataset.

\end{itemize}

\subsection{Anisoplanatism}

All existing AO systems suffer from \emph{anisoplanatism}, meaning that the shape of a PSF is a function of its position in the field of view. Usually, this spatial variability is described by the relation

\begin{equation}
P(\theta) = P(0) * K(\theta)
\label{eq_aniso_convol}
\end{equation}

\noindent
(Fusco et al. \cite{fusco2000}; Steinbring et al. \cite{steinbring2002,steinbring2005}; Cresci et al. \cite{cresci2005}). Here $\theta$ is the angular distance from the center of AO correction, $P(0)~ [P(\theta)]$ is the PSF at the center of correction [at distance $\theta$], and $K(\theta)$ is a kernel describing the PSF variation; $*$ denotes convolution.

\begin{figure}
\includegraphics[width=8.8cm]{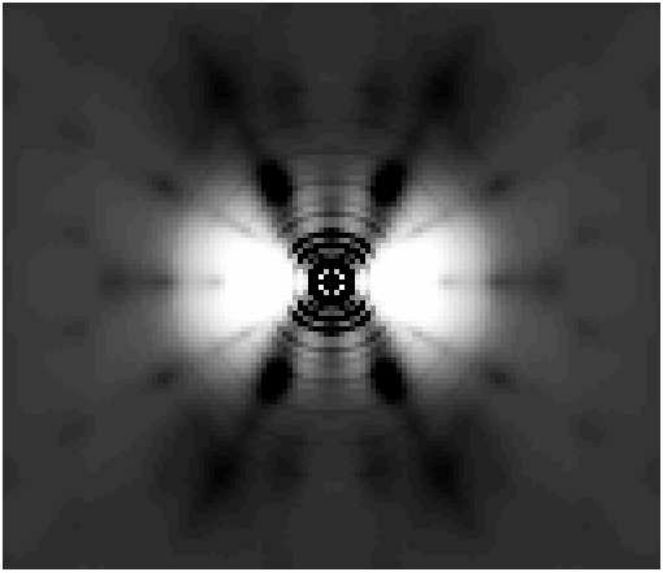}
\caption{Difference image of two simulated MAORY K-band PSFs located 70'' away from each other at opposite sides of the center of AO correction. This image covers a field of $\approx7.5''\times6''$. The difference map shows a complicated, highly symmetric pattern.}
\label{fig_kernel}
\end{figure}

If $K(\theta)$ were asymmetric, then the center of light of the PSF -- and thus the detector position -- would be a function of $\theta$. This would introduce a systematic distortion of astrometric solutions that has to be calibrated out.

For existing SCAO systems, $K(\theta)$ is known to be symmetric. It can be approximated analytically as an elliptical Gaussian profile (e.g. Steinbring et al. \cite{steinbring2005}). In this case, the most important impact of anisoplanatism is a systematic degradation of the PSF's Strehl ratio $S$ like

\begin{equation}
S(\theta) = S_0\exp [-(\theta/\theta_N)^{5/3}]
\label{eq_aniso_strehl}
\end{equation}

\noindent
assuming a perfect AO correction. Here $S_0$ is the Strehl ratio at the center of AO correction, the parameter $\theta_N$ is the anisopanatic angle. This effect reduces the SNR of a source and thus the statistical astrometric accuracy, but a priori does not introduce a systematic error.

\begin{figure}
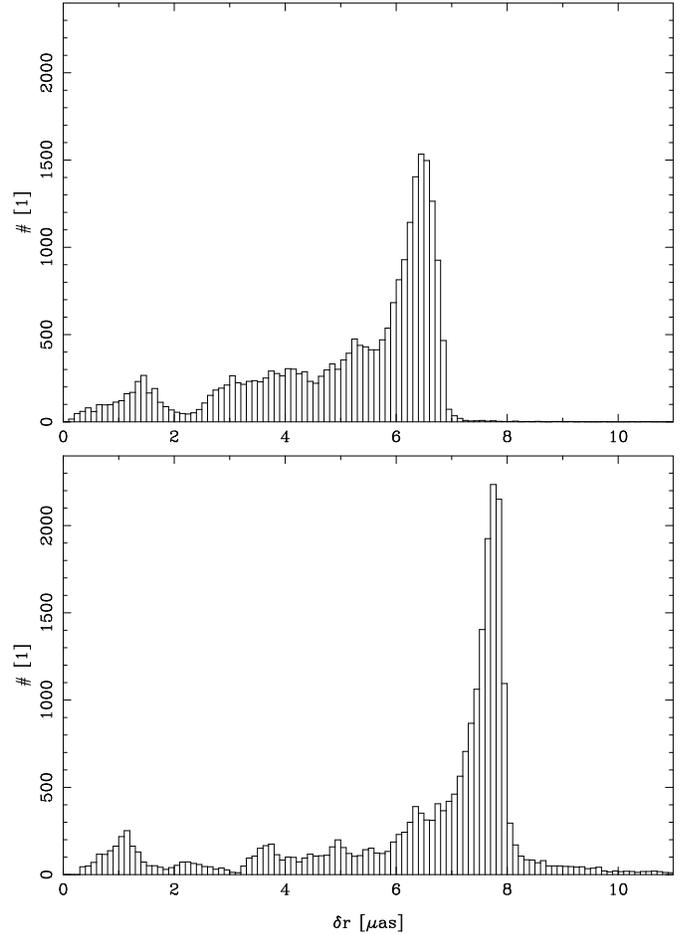

\includegraphics[height=8.8cm,angle=-90,trim = 0cm 0cm 1cm 0cm, clip]{aniso_histo_H.eps}
\includegraphics[height=8.8cm,angle=-90]{aniso_histo_K.eps}
\caption{Astrometric errors introduced by MCAO anisoplanatism in H (\emph{top panel}) and K (\emph{bottom panel}) bands. Errors are centroid shifts (with respect to the case of perfectly symmetric light distributions) of the difference images between pairs of simulated PSFs located at different locations in the MAORY FOV. The histograms have cutoffs at $\approx$7$\mu$as in H band and at $\approx$8$\mu$as in K band.}
\label{fig_aniso}
\end{figure}

For the case of MCAO systems with multiple centers of correction the problem is less well understood. In order to estimate the astrometric error introduced by anisoplanatism, we made use of preliminary simulations of multi-guide-star MAORY PSFs in H and K bands\footnote{Made available by the MAORY consortium at {\tt http://www.bo.astro.it/$\sim$maory/Maory/}}. The simulation data provide a grid of 217 PSFs located at distances between 0'' and 80'' from the center of correction. The simulations assume a seeing of 0.6''. PSFs are sampled with pixel scales equal to $\lambda/2D$, corresponding to 4.05~mas/pix and 5.30~mas/pix for H and K bands, respectively. 

By construction, each PSF is centered accurately at the central pixel of a 512$\times$512 pixel grid. As discussed above, we need to estimate the relative position error introduced by anisoplanatism. For this we computed for each pair of PSFs a difference map by subtracting one PSF image from the other. In Fig.~\ref{fig_kernel} we show the difference image for the case of two PSFs located 70'' away from each other at positions $(-60'',0'')$ and $(+10'',0'')$ with the center of AO correction located at $(0,0)$. Clearly, the map shows a complicated, highly symmetric light distribution. From the high degree of symmetry one can already suspect qualitatively that the impact of MCAO anisoplanatism on the astrometry is small.

In order to quantify this statement, we calculated for each difference map the centroid

\begin{equation}
x_{\rm centroid} = \frac{\sum_{i}x_i\cdot I_i}{\sum_{i}I_i}
\label{eq_centroid}
\end{equation}

\noindent
of the light distribution. Here $x$ is an arbitrary coordinate, $x_i$ and $I_i$ are position and flux assigned to the $i$-th pixel. The summation is performed over all map pixels (e.g. Berry \& Burnell \cite{berry2000}). As we had 217 PSF images at hand, this procedure resulted in 23436 measurements. Centroid positions different from the center of the pixel grid correspond to position shifts of one PSF with respect to the other caused by asymmetries in the anisoplanatism kernel.

We summarize our results in Fig.~\ref{fig_aniso}. The histograms of the position shifts show quite sharp cutoffs at $\approx$7$\mu$as in H band and at $\approx$8$\mu$as in K band. We do not find a significant correlation of position errors with distance from the center of correction or any other systematic relation.

We conclude that MCAO anisoplanatism probably introduces very small but noticable uncertainties into astrometry. Depending on wavelength band we find errors (upper ends of histograms) up to $\approx$8$\mu$as. As the anisoplanatism effect is not (or very weakly) correlated with PSF positions in the FOV, it can probably not be caught by coordinate transforms or any other systematic correction but has to be included into the error budget.

\subsection{High-$z$ Galaxies As Reference Sources}

Throughout this article we discuss \emph{relative} astrometry, meaning measuring relative source positions. So far we implicitly assumed the reference sources to be point sources, especially stars. However, in some cases the use of reference points other than stars will be necessary:

\begin{itemize}

\item  The number of stars located in the FOV can be too small. As discussed in Sect.~3, the number of reference sources is a function of the order of the coordinate transform used. For example, in case of a 2nd-order transform, the minimum number is six. Using a larger number than the theoretical minimum is useful to average out statistical position errors that could propagate into the transformation. Additionally, systematic effects like instrumental distortion (Sect.~4.2) and differential tilt jitter (Sect.~4.7) require large numbers of reference points for calibration purposes. Deep fields at moderate galactic latitudes ($|b|\approx50^{\circ}$) include only a few to a few tens $K_{\rm AB}<$23 stars/arcmin (see e.g. F\"orster Schreiber et al. \cite{foerster2006}; Wuyts et al. \cite{wuyts2008}). This number is probably too small for astrometric MICADO observations.

\item  The science case might require an extragalactic reference frame, thus excluding Galactic stars. For example, several groups (e.g. Bedin et al. \cite{bedin2003,bedin2006}; Kalirai et al. \cite{kalirai2004}) use samples of background galaxies as reference in order to measure the absolute motion of globular clusters. Other science cases may need similar approaches (see Sect.~2).

\end{itemize}

In order to investigate the use of galaxies as references, we analyzed two simulated MICADO deep field K-band images astrometrically. We simulated galaxies following realistic distributions of K-band magnitude as function of redshift $z$ taken from the Chandra Deep Field South FIREWORKS (``field A''; Wuyts et al. \cite{wuyts2008}) and FIRES Hubble Deep Field South (``field B''; Labb\'e et al. \cite{labbe2003}) catalogs. All objects in the fields are built using four main ingredients:

\begin{itemize}

\item  Smooth elliptical and disk galaxy light distributions with realistic surface number densities, sizes, and K-$z$ distributions.  The  K-$z$  distributions were obtained by drawing at random galaxies from the two source catalogues above, scaling the numbers by the ratio of the MICADO FOV and the respective survey areas.  A fixed mix of ellipticals and disks was adopted (40\%/60\%), and effective radii were assigned at random so as to match approximately the size distributions of Franx et al. \cite{franx2008}. Uniform distributions are assumed for the axis ratios (with minimum of 0.3 for ellipticals and 0.1 for disks) and for the position angles.

\item  Bulges (for disk galaxies) with de Vaucouleurs luminosity profiles adopted for simplicity.  Bulge properties are poorly constrained at high redshift; our assumptions were guided by results from Elmegreen et al. (\cite{elmegreen2009}, and references therein; see also Genzel et al. \cite{genzel2008}).  We assumed uniform distribution of axis ratios between 0.7 and 1, position angles fixed at those of the host disks, effective radii inversely proportional to $1+z$ and such that $r_e=2$kpc at $z=0$, and a random distribution of bulge-to-total light ratios between 0 and 0.5.

\item  Clumps (for disk galaxies) with realistic numbers per disk, typical light fractions, characteristic sizes and dependence on redshift, and radial distribution across the disks so as to roughly match observed properties at $z\approx$1--2.5 (e.g., Genzel et al. \cite{genzel2008}; Elmegreen et al. \cite{elmegreen2009} and references therein) and their lack at $z\approx0$. Of order 1--10 clumps per disk were simulated, with Gaussian light profiles (again for simplicity), a narrow distribution of FWHMs varying as $(1+z)$ such that FWHM = 1~kpc at $z=2.5$, light fractions of a few percent typically, an exponential distribution of radial positions with scale-length five times that of the host disk, random position angles and axis ratios between 0.7 and 1.

\item  Unresolved star clusters (for all galaxies).  This is the ingredient which is least constrained at high redshift, as current instruments lack the resolution and point-source sensitivity to detect individual clusters if such exist at high $z$.  To define their properties, we assumed superpositions of five different populations of point-like sources with a range of (Gaussian) magnitude and radial distributions. The brightest population contains the fewest clusters, with the most centrally concentrated distributions.  The brightest cluster in each galaxy contributes 0.1\% of the total light. These assumptions roughly reproduce those of super star clusters and globular clusters in local galaxies, and unresolved sources in local analogs of z ~ 3 Lyman-break galaxies (Overzier et al. \cite{overzier2008}).

\end{itemize}

\begin{figure}
\includegraphics[width=4.3cm]{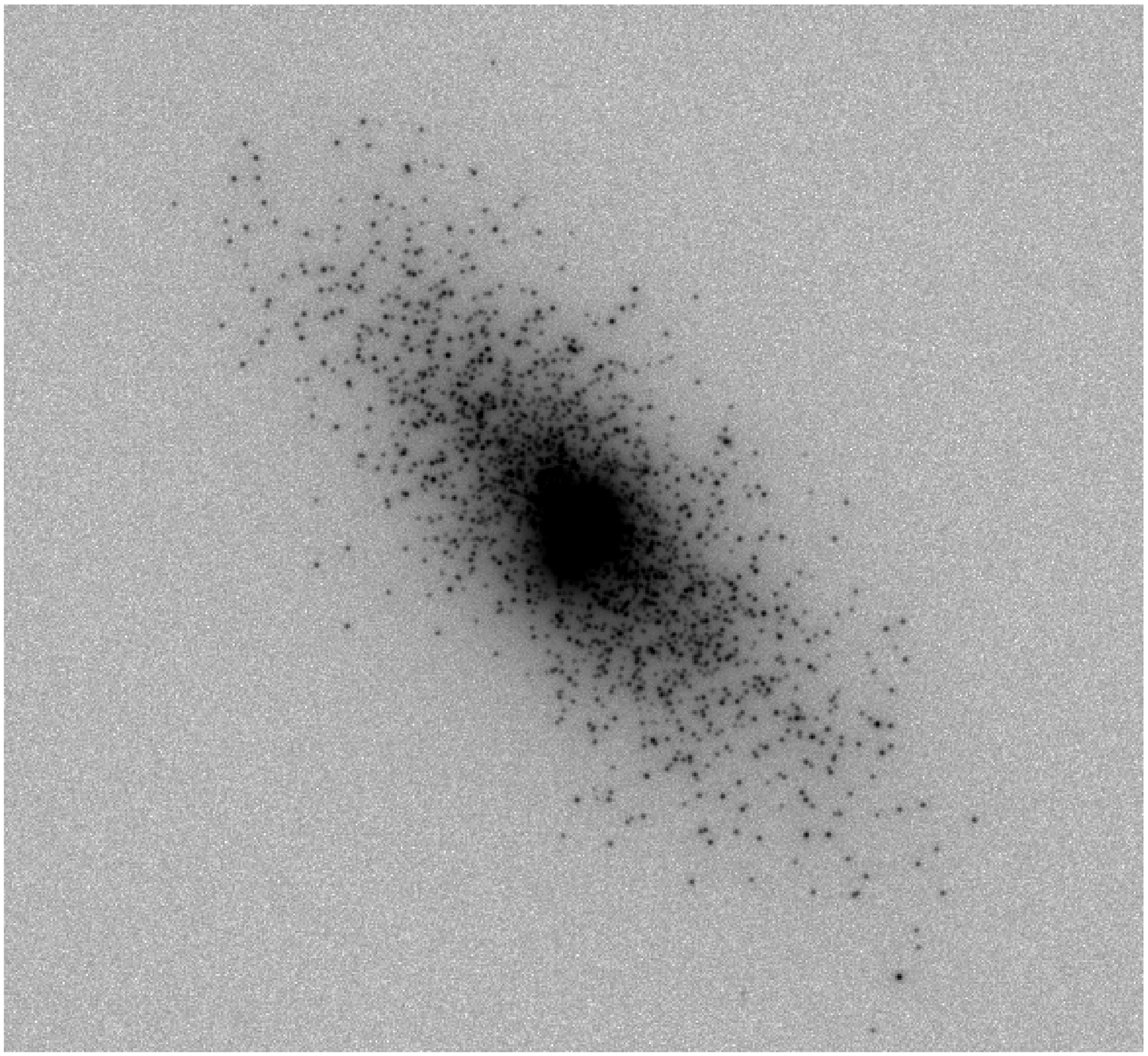}
\includegraphics[width=4.3cm]{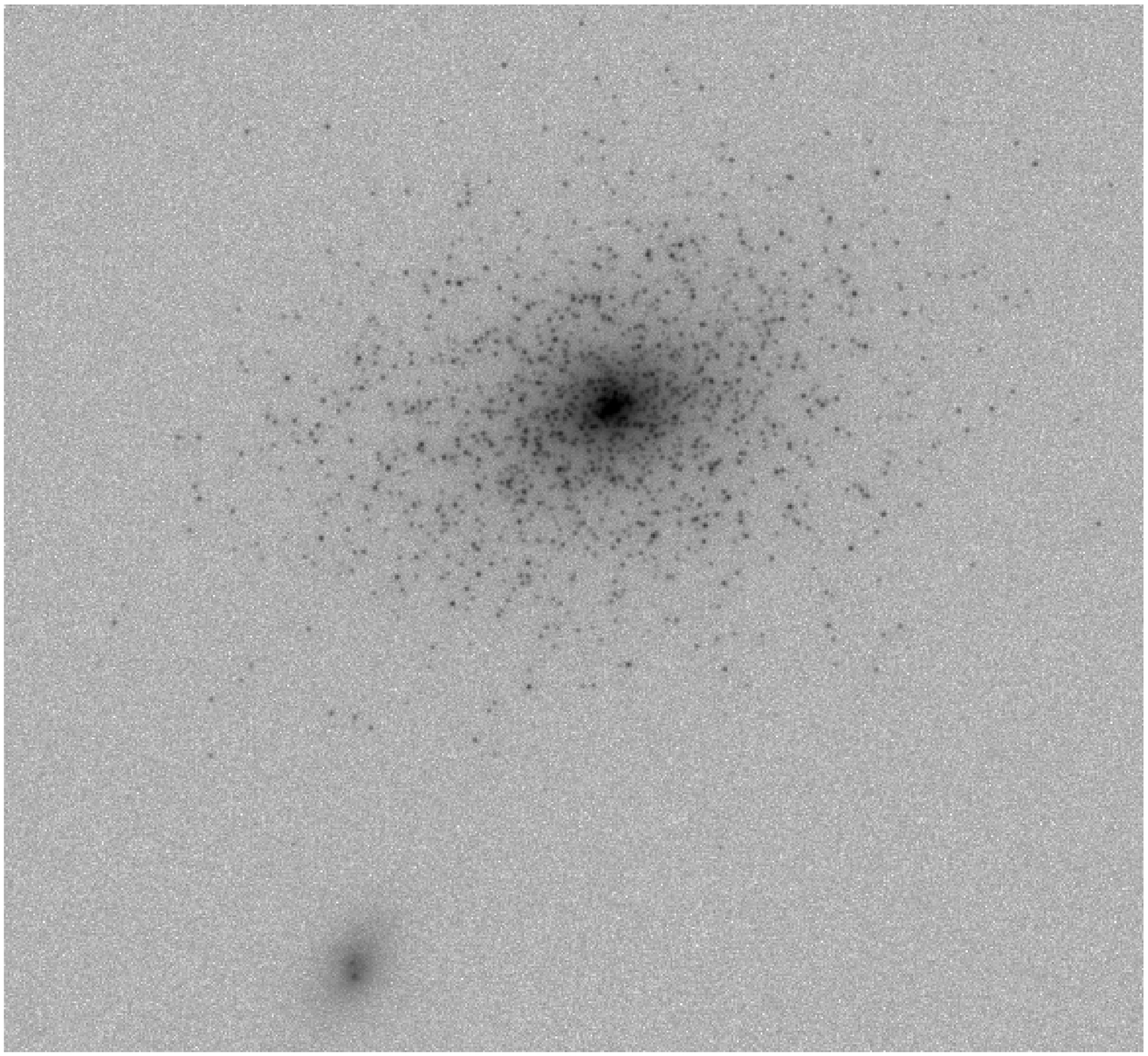}

\includegraphics[width=4.3cm]{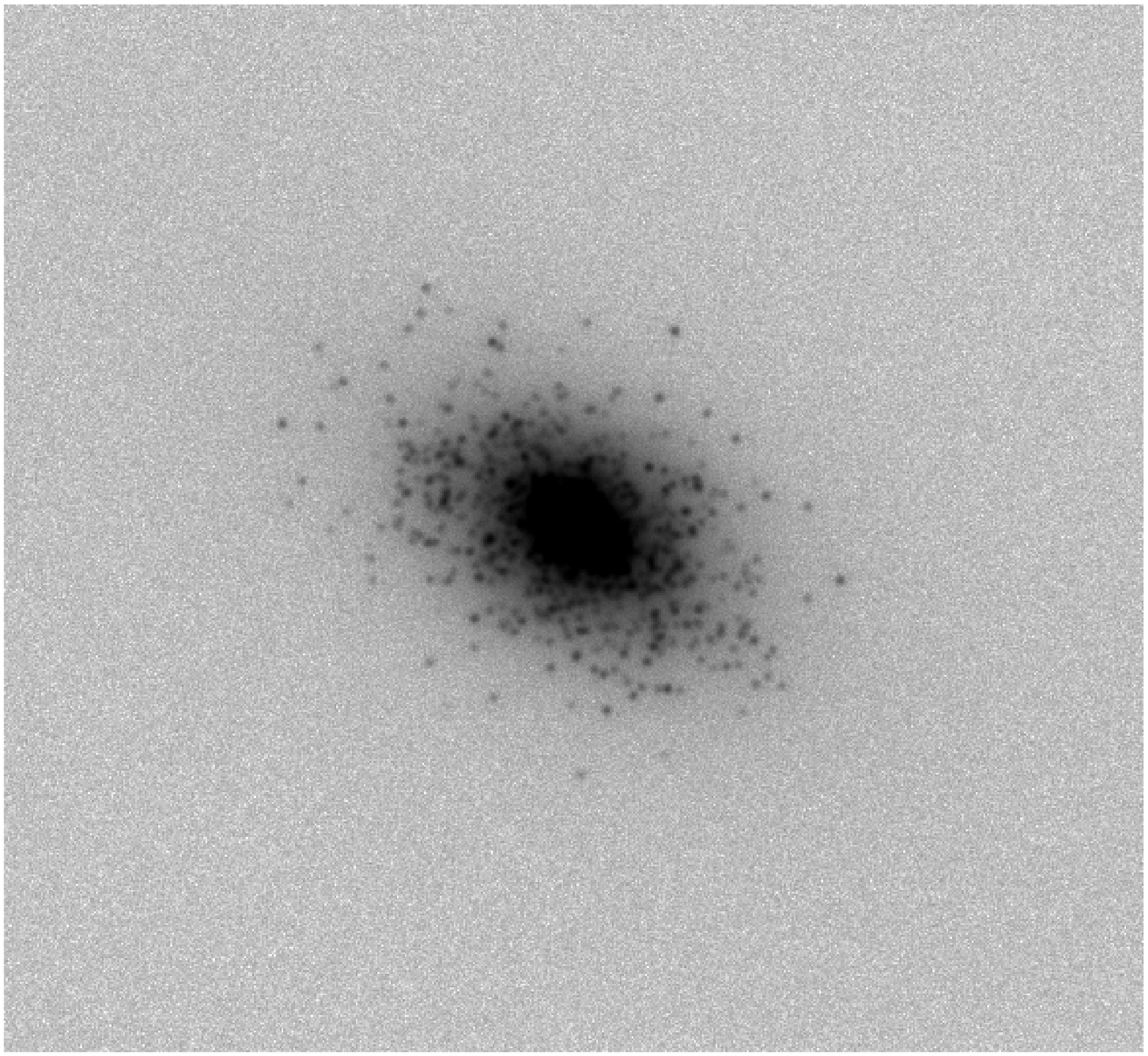}
\includegraphics[width=4.3cm]{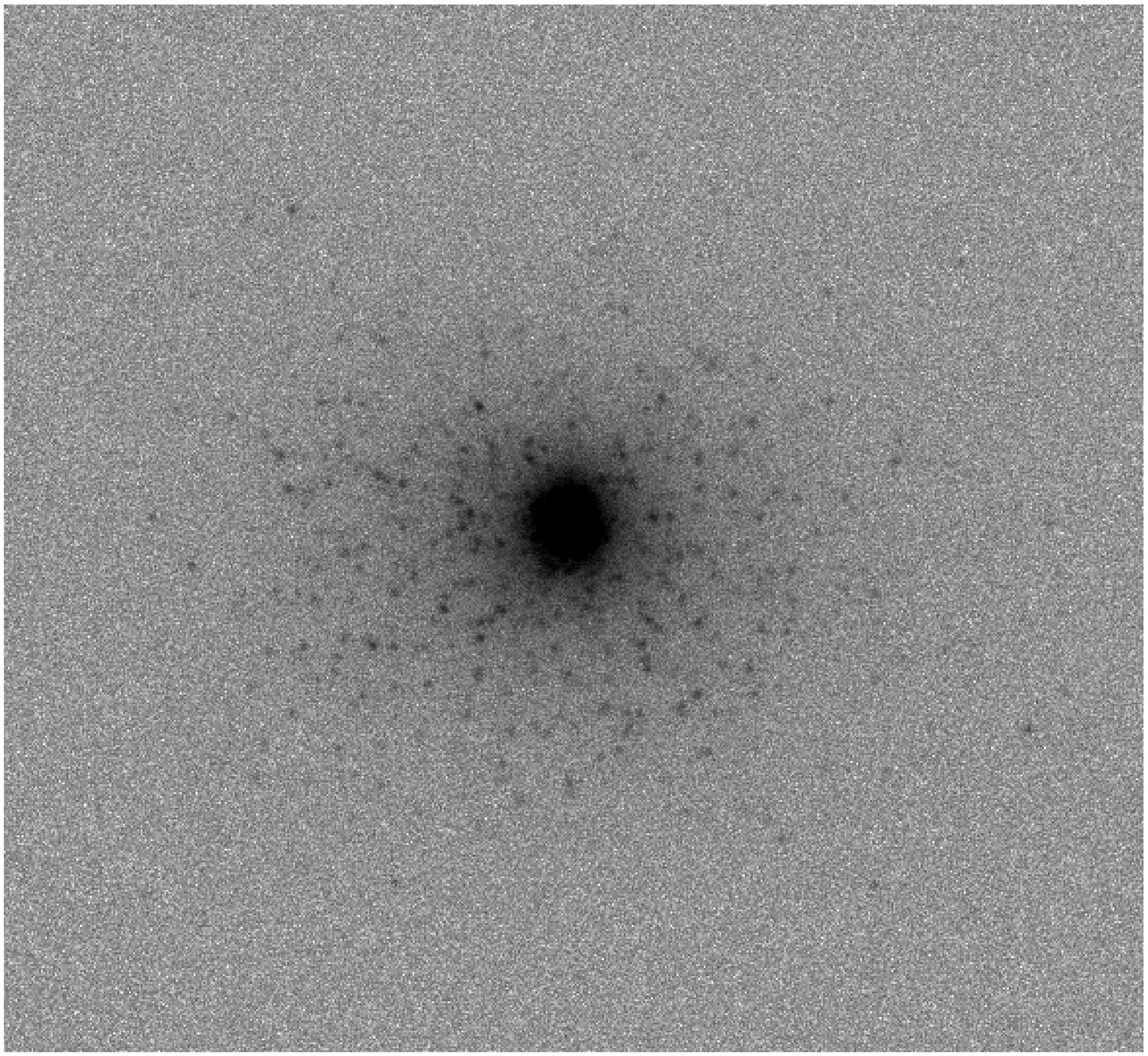}
\caption{A few simulated galaxies from our artificial deep field B. Images in the top line cover $\approx3.5''\times3.5''$, images in the bottom line cover $\approx1.8''\times1.8''$.}
\label{fig_highz_sim}
\end{figure}

\noindent
For each field image we created three realizations corresponding to E-ELT/MICADO integration times of 1, 4, and 10 hours, respectively, by adding Gaussian noise to the original map. As given by the input catalogs, field A has a point source brightness limit of $K_{\rm AB}=24.3$, field B of $K_{\rm AB}=25.6$. These limits are $5\sigma$ depths for point sources in circular apertures of 2'' diameter. In Fig.~\ref{fig_highz_sim} we show four simulated galaxies from field B as examples.

The star clusters are especially interesting for astrometry. They provide an ensemble of several thousands of bright, point-like sources that can be analyzed with standard tools developed for point source astrometry. In each field we selected all star clusters that were sufficiently isolated, meaning a distance of at least 30~mas (about three resolution elements) to the next source. In total we used 1600 sources in field A and 2174 sources in field B. In each image we fit the clusters with 2D Gaussian brightness distributions in order to accurately (meaning few milli-pixels in the best cases) determine their positions. We then compared the results with the true source positions on the original noise-free maps. The distributions of the deviations between true and measured positions provide measures of astrometric accuracies.

\begin{figure}
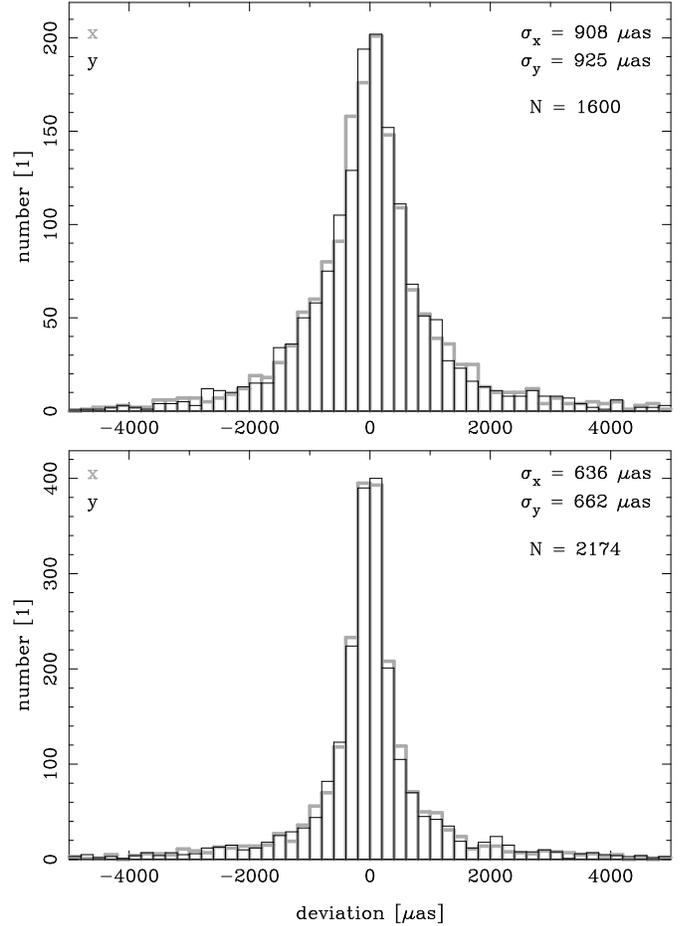

\includegraphics[height=8.8cm,angle=-90,trim = 0cm 0cm 1cm 0cm, clip]{devhisto_v2.eps}
\includegraphics[height=8.8cm,angle=-90]{devhisto_v4.eps}
\caption{Histograms of position errors found from using several thousand point-like star clusters as references. These results are for fields A (top panel) and B (bottom panel) at 10h integration time. Please note the different number axes scales.}
\label{fig_highz_err}
\end{figure}

We show the results for 10h integration times in Fig.~\ref{fig_highz_err}. For both fields we find distributions with 1$\sigma$-widths of $\approx$920$\mu$as (field A) and $\approx$650$\mu$as (field B) per coordinate. These numbers correspond to the typical measurement accuracies for individual star clusters. The global astrometric accuracy is given by the standard error of the mean of the distribution, i.e. $\sigma/\sqrt{N}$ with $N$ being the number of sources.

We summarize the astrometric accuracies found for all fields and integration times in Table~\ref{tab_highz}. Not surprising, we find the highest accuracies of 23$\mu$as and 14$\mu$as (per coordinate) for fields A and B, respectively, for the largest integration time of 10h. The errors increase to about 65$\mu$as and 29$\mu$as at integration times of 1h. The systematic difference between fields A and B originates from their different depths: field B is based on a catalog that has a $\approx$1.3 magnitudes deeper brightness limit; it is therefore ``richer'' of sufficiently bright targets. As MICADO will perform astrometric observations of sources $K_{\rm AB}<26$ (see also Sect.~1), the results found from field B are actually more realistic for describing the performance of MICADO. We can therefore conclude that total (i.e. the quadratic sum of the values for both coordinates) astrometric accuracies of $\approx$20$\mu$as are realistic for integration times of order 10h when using galactic star clusters as referencess.

So far, we have taken into account non-resolved (point-like) sources as references only. However, the global extended light distribution contains additional information. We therefore computed 2D cross-correlations between different realizations of the same ``observations''. As reference we used one 7000$\times$7000 pixels (21''$\times$21'') sub-field of field A containing seven galaxies. For each integration time (1h, 4h, 10h) we computed three realizations of the random noise map added to the original light distribution. For all pairwise combinations we computed the 2D cross-correlations. We calculated their centers by means of 2D Gaussian fits to the central parts of the cross-correlation maps. The uncertainty of the map centers provides a measure of astrometric accuracy. For all integration times we found very similar results: 4.5$\mu$as at 1h, 4.1$\mu$as at 4h, and 3.9$\mu$as at 10h. This indicates that at these level the accuracy is only weakly correlated with the SNR but dominated by systematic effects. In any case the impressive accuracies of $\approx$4$\mu$as suggest that using 2D cross-correlations of images can provide highly accurate astrometry.

However, 2D cross-correlations cannot be used for relative astrometry in a straight forward manner. By construction, cross-correlations are only sensitive to shifts between images. They cannot describe transformations like scalings, rotations, or higher order transforms as discussed in Sect.~3. Using them for astrometry therefore requires a more sophisticated approach than computing the cross-correlation between two (or more) images. One ansatz might be to calculate cross-correlations for a sufficient number (about ten or more) of sub-fields of the target area. This would provide a set of relative position (changes). Feeding this dataset into a proper model might result in a highly accurate coordinate transform.

\begin{table} 
\centering
\caption{Astrometric accuracies found from using several thousand point-like star clusters as references. We give errors for the coordinates $x$ and $y$ as function of field and integration time.}
\begin{tabular}{c c c c c}
\hline\hline
 & field A &  & field B &  \\
\hline
$t_{\rm int}$ [h] & $\delta x$ [$\mu$as] & $\delta y$ [$\mu$as] & $\delta x$ [$\mu$as] & $\delta y$ [$\mu$as] \\
\hline
10 & 23 & 23 & 14 & 14 \\
\hline
4 & 30 & 36 & 17 & 17 \\
\hline
1 & 58 & 65 & 29 & 29 \\
\hline
\end{tabular}
\label{tab_highz}
\end{table}

From our analysis we conclude that the use of high-$z$ galaxies as astrometric calibrators is feasible at the level of $\approx$20$\mu$as at integration times of about 10h. This number assumes the use of several thousand point-like star clusters as reference sources (also assuming that such exist at intermediate to high redshift). Cross-correlation of images might provide even better accuracies, but would require a sophisticated computational scheme that can actually provide full coordinate transforms.

In some cases, one might consider separating the intra-epoch and the inter-epoch calibration steps. Firstly, one can use a small number of stars (as discussed earlier, six may be sufficient) for \emph{intra}-epoch (frame-to-frame) calibration. Proper adding of images allows building up SNR sufficient for detection of and position measurements on faint galaxies. Secondly, one can use these galaxies for \emph{inter}-epoch calibration.

\subsection{Calibration of the Projected Pixel Scale}

In Section~3 we have outlined the concepts of relative astrometry and coordinate transforms. These schemes make use of sets of reference positions $\{{\bf x}_{\rm ref}^n\}$ in order to compute transformation matrices for the $n$-th dataset. However, all transformation and calibration steps we have discussed up to now are executed in image space, i.e. in units of pixels. After removing non-linear distortions from the data, one needs to calibrate the pixel scale \emph{as projected on sky} in order to accurately convert measured positions and motions into angular units. We note that the projected pixel scales for the two coordinates $x, y$ may be different if the detector plane is tilted with respect to the focal plane. 

Calculating the scaling factors requires astrometric reference points with known positions located in the target FOV. The number of reference sources should be at least three in order to allow for a full linear transformation. Fortunately, this calibration step needs to be executed only once for a selected ``master'' (or ``zero'') image. All other images can be connected to the master image reference frame via coordinate transforms, including the proper scaling (e.g. Trippe et al. \cite{trippe2008b}). Errors on the reference positions propagate into the positions and motions calculated from the data. For the FOV of MICADO, reference position accuracies of order 1~mas translate into a \emph{relative} scaling accuracy of

$$
\delta x / x \approx 1 {\rm mas} / 53'' \approx 2\times10^{-5}
$$

As discussed in Section~2, the science cases for E-ELT/MICADO demand accurate measurements of positions and proper motions over spatial scales from few ten $\mu$as (e.g. parallaxes of globular clusters; see Sect.~2.3) to few hundred mas (e.g. stellar orbits around Sgr~A*; see Sect.~2.1). A relative pixel scale accuracy of $2\times10^{-5}$ corresponds to an error of 10~$\mu$as over an angular distance of 500~mas. From this we take the message that the accuracies of the reference positions should not exceed $\approx$1~mas substantially. 

There are several possibilities for obtaining very accurate reference source positions. One of them is the use of sources located in the MICADO FOV and visible in both NIR and radio, e.g. QSOs or maser stars. If radio-interferometric positions (e.g. VLBI) -- which are accurate on the sub-mas level -- are at hand, the uncertainty on the pixel scale can be very low (see, e.g., Reid et al. \cite{reid2007} for the case of the Galactic center). Another option might be the use of stars from the HIPPARCOS catalogue (Perryman et al. \cite{perryman1997}). However, the uncertainties in stellar proper motions, which are $\approx$0.8~mas/yr, will propagate into position uncertainties of $\approx$20~mas over a timeline of 25 years (given that the HIPPARCOS catalogue reference epoch is J1991.25; Perryman et al. \cite{perryman1997}). This means that the astrometric uncertainties will exceed 10~$\mu$as if the spatial scale of the experiment is larger than $\approx$25~mas. A more promising approach is the use of data from the future GAIA astrometry space mission that will provide absolute astrometric accuracies well below 1~mas (Jordan \cite{jordan2008}). The main advantage of the GAIA catalogue (its high accuracies aside) is the large number of stars included -- $\approx$10$^9$ -- meaning that for any arbitrary field in the sky a sufficient number of reference points should be available.

We conclude from our discussion that for MICADO sufficient numbers of calibration sources with absolute position accuracies better than 1~mas will be available. We therefore expect that the calibration of the projected pixel scale introduces errors of $\approx$10$\mu$as at most.

\section{Results and Error Budget}

We have identified and discussed ten effects that might limit the expected astrometric accuracy of E-ELT/MICADO observations systematically. We have been able to quantify each of these sources of error. From this, we can calculate a prediction for the error budget of MICADO.

\begin{enumerate}

\item  For isolated point sources, \emph{detector sampling / binning} does not introduce noticable ($\approx$1$\mu$as) errors as long as the pixel scale does not exceed 3~mas/pix. For sources affected by crowding, using a smaller scale of 1.5~mas/pix can improve the accuracy by factors of about two compared to the 3~mas/pix case. Therefore MICADO will use a pixel scale of 3~mas/pix as standard and a reduced scale of 1.5~mas/pix for mapping crowded fields. For the error budget, we can thus assume a sampling error

$$
\sigma_{\rm samp} = 1~\mu\rm as~.
$$

\item  Instrumental \emph{geometric distortion} needs to be taken into account by dedicated calibration procedures. We propose to implement a calibration mask into MICADO that illuminates the detectors with a well-defined image. Such a mask would have to be mapped with accuracies of $\approx$40nm. Based on our results in combination with published works using Hubble Space Telescope data, we estimate that distortion can be corrected down to levels of $\approx$10--30$\mu$as. For the error budget, we therefore use

$$
\sigma_{\rm dist} = 30~\mu\rm as~.
$$

\item  \emph{Telescope instabilities}, notably plate scale instabilities and instrumental rotations, are linear effects that can be absorbed by coordinate transforms. Therefore they do not contribute to the error budget.

\item  Atmospheric \emph{achromatic differential refraction} is important (order 10~mas) only in linear terms which can absorbed by coordinate tranforms. Higher-order contributions are of order 1$\mu$as, meaning for the error budget

$$
\sigma_{\rm ADR} = 1~\mu\rm as~.
$$

\item  \emph{Chromatic differential refraction} introduces position errors of the order $\approx$1~mas in NIR observations depending on relative source colours. A tuneable ZnS/ZnSe atmospheric dispersion corrector can reduce this effect to $\approx$10--20$\mu$as for ``typical'' science cases. Extreme relative source colours might require the additional use of narrowband filters and/or analytic a posteriori correction schemes. For the error budget, we thus set

$$
\sigma_{\rm CDR} = 20~\mu\rm as~.
$$

\item  AO \emph{guide star measurement errors} for N natural guide stars can introduce distortions up to order N$-$1 into images, meaning 2nd order distortions for the three NGS of MICADO. This effect can be absorbed by coordinate transforms (of order N$-$1). It therefore does not contribute to the error budget.

\item  Atmospheric \emph{differential tilt jitter} can introduce errors of $\approx$100$\mu$as into diffraction-limited E-ELT observations. It integrates out with $t^{-1/2}$. For MICADO which uses an MCAO system, the tilt jitter error can be integrated down to $\approx$10$\mu$as within about 30 minutes of observation. Using dedicated coordinate transform allows reaching this accuracy in shorter times. For the error budget, we thus set

$$
\sigma_{\rm TJ} = 10~\mu\rm as~.
$$

\item  The \emph{anisoplanatism} of the MAORY AO system introduces uncertainties of up to $\approx$8$\mu$as. There appears to be no correlation of this error with the position of a PSF in the FOV; it therefore can not be calibrated out in a straight forward manner. Therefore we add it to the error budget:

$$
\sigma_{\rm aniso} = 8~\mu\rm as~.
$$

\item  Depending on target field and science case, the use of \emph{galaxies as astrometric calibrators} may be necessary. From a simulated MICADO deep field we find that we can use several thousand non-resolved galactic star clusters as point-like reference sources. However, good accuracies of $\approx$20$\mu$as require long intra-epoch integration times of about 10 hours; we consider this to be a somewhat large but realistic timescale. We therefore use for the error budget

$$
\sigma_{\rm galaxies} = 20~\mu\rm as~.
$$

\item  The accuracy of the \emph{sky-projected pixel scale} is limited by the accuracy of astrometric positions of reference sources in the FOV. Given the typical accuracies of present-day radiointerferometric data and the future GAIA catalogue, which are better than $\approx$1~mas, we add to the error budget

$$
\sigma_{\rm scale} = 10~\mu\rm as~.
$$

\end{enumerate}

From the individual uncertainties listed above we can calculate a total intrinsic astrometric accuracy for MICADO as

\begin{equation}
\sigma_{\rm sys} = \sqrt{\sum_i \sigma_i^2} = 44\mu\rm as ~ . 
\label{eq_errbudget}
\end{equation}

\noindent
This number provides a systematic limit for astrometric accuracies to be expected from MICADO data. Of course, this result corresponds to a somewhat arbitrary ``typical'' case. As many parameters like integration times, source colours, numbers and types of reference sources, etc. can vary over wide ranges, the actual $\sigma_{\rm sys}$ for a specific observation can be quite different -- in both directions -- from the one we quote here. Nevertheless, we conclude that we are able to quantify the mean systematic astrometric accuracy achievable with MICADO which is $\approx$40$\mu$as.

When discussing the accuracy of the measurement for a given target, one of course needs to add the statistical measurement error $\sigma_{\rm L}$ (Eq.~\ref{eq_accuracy}) which scales with the SNR. For SNR$=$100, $\sigma_{\rm L}=34\mu$as, and thus the combined error is $\sqrt{\sigma^2_{\rm sys} + \sigma^2_{\rm L}} = 56\mu\rm as$. SNRs different from 100 modify this number accordingly.

\section{Conclusions}

In this article we have studied the capabilities expected for the NIR imager MICADO for the future 42-m European Extremely Large Telescope with respect to accurate astrometry. A variety of science cases requires long-term astrometric accuracies of $\approx$50$\mu$as. We discuss and quantify ten effects that potentially limit the astrometric accuracy of MICADO. We conclude that the systematic accuracy limit for astrometric observations with MICADO is $\sigma_{\rm sys}\approx40\mu$as. We find that astrometry at this accuracy level with MICADO requires the fulfillment of several conditions:

\begin{itemize}

\item  All images, regardless of their distance in time, need to be combined via full coordinate transforms of second order or higher.

\item  MICADO needs to be equipped with an astrometric calibration mask for monitoring the instrumental distortion. The pixel scale of the camera should not exceed the 3~mas/pix used in the current design.

\item  Astrometric observations require decent integration times of at least 30 minutes per epoch. This is unavoidable in order to average out atmospheric tilt jitter. When using high-z galaxies as astrometric reference points, integration times up to about 10 hours can be necessary.

\end{itemize}

It is noteworthy that the effects discussed in this article already affect observations collected with present 8m-class telescopes. In his exhaustive analysis of NIR images obtained with VLT/NACO, Fritz \cite{fritz2009} has been able to detect signatures of chromatic differential refraction and differential tilt jitter in his astrometric dataset. He concludes that taking into account these effects can improve the accuracies down to few hundred $\mu$as. This agrees with the findings of Lazorenko \cite{lazorenko2006} and Lazorenko et al. \cite{lazorenko2007} who analyze seeing-limited optical R-band ($\lambda_{\rm center}=655$nm) images taken with VLT/FORS1+2. They conclude that they are able to achieve astrometric precisions (but not accuracies) of $\approx$100$\mu$as by using a special scheme for scheduling observations and dedicated coordinate transforms (although they neglect instrumental geometric distortion).

The analysis we provide here is set up for the specific case of E-ELT/MICADO, but parts of our results are valid in general. This study should thus contain valuable information for other future 30--40m telescopes. As some of the effects we discuss are actually observed in present day 8m-class telescope data, our analysis might also be helpful for the calibration of data already taken. We therefore expect that our work is of interest well beyond the E-ELT community.

\section*{Acknowledgements}

We are grateful to the members of the MAORY consortium for providing us with simulated MCAO PSFs. N.M.F.S. gratefully acknowledges support from the Minerva Program of the Max-Planck-Gesellschaft. Last but not least, we thank the anonymous reviewer whose comments were helpful to improve the quality of this article.

\label{lastpage}

\end{document}